\documentclass[apj]{emulateapj}

\newcommand  \acc     {\ifmmode {\rm km\,s}^{-2} \else km\,s$^{-2}$\fi}

\newcommand  \ergs     {\ifmmode {\rm ergs\,s}^{-1} \else ergs s$^{-1}$\fi}
\newcommand  \ergcms   {\ifmmode {\rm erg\,cm}^{-2}\,{\rm s}^{-1}
                        \else ergs\,cm$^{-2}$\,s$^{-1}$\fi}
\newcommand  \ergcmsA  {\ifmmode{\rm erg\,cm}^{-2}\,{\rm s}^{-1}\,{\rm\AA}^{-1}
                        \else ergs\,cm$^{-2}$\,s$^{-1}$\,\AA$^{-1}$\fi}
\newcommand  \ergcmsHz {\ifmmode{\rm ergs\,cm}^{-2}\,{\rm s}^{-1}\,{\rm Hz}^{-1}
                        \else ergs\,cm$^{-2}$\,s$^{-1}$\,Hz$^{-1}$\fi}
\newcommand  \phcms    {\ifmmode {\rm ph\,cm}^{-2}\,{\rm s}^{-1}
                        \else ph\,cm$^{-2}$\,s$^{-1}$\fi}
\newcommand  \phcmsA   {\ifmmode {\rm ph\,cm}^{-2}\,{\rm s}^{-1}\,{\rm\AA}^{-1}
                        \else ph\,cm$^{-2}$\,s$^{-1}$\,\AA$^{-1}$\fi}

\slugcomment{Received 2004 December 22; accepted 2005 February 16}

\shorttitle{UV VARIABILITY IN LINERS}

\shortauthors{MAOZ ET Al.}

\begin{document}

\title{
The Murmur of the Sleeping Black Hole:
Detection of Nuclear Ultraviolet Variability in LINER Galaxies
\footnotemark[1]} \footnotetext[1]{
Based on observations with the {\it Hubble Space
Telescope} which is operated by AURA, Inc., under NASA contract NAS
5-26555.}
\author{
Dan~Maoz,\altaffilmark{2} 
Neil M. Nagar, \altaffilmark{3,4}
Heino Falcke, \altaffilmark{5,6} 
and Andrew S. Wilson, \altaffilmark{7}  
}

\altaffiltext{2}{School of Physics and Astronomy, 
Tel-Aviv University, Tel-Aviv 69978,
Israel; dani@wise.tau.ac.il}
\altaffiltext{3}{Kapteyn Institute, Postbus 800, 9700AV, Groningen, 
The Netherlands}
\altaffiltext{4}{Departamento de F\'isica, Astronomy Group, 
Universidad de Concepci\'on,
  Casilla 160-C, Concepci\'on, Chile}
\altaffiltext{5}{ASTRON, P.O. Box 2, 7990 AA, Dwingeloo, The
  Netherlands}
\altaffiltext{6}{Department of Astrophysics, Radboud University, 
Postbus 9010, 6500 GL
Nijmegen, The Netherlands}
\altaffiltext{7}{Astronomy Department, University of Maryland, College Park, 
MD 20742}

\begin{abstract}
LINER nuclei, which are present in many nearby galactic bulges,
may be the manifestation of low-rate or low-radiative-efficiency 
accretion onto 
supermassive central black holes. However, it has been unclear whether 
the compact ultraviolet (UV) nuclear sources present in
many LINERs are clusters of massive stars, rather than being 
directly related to the accretion process.   
We have used the Hubble Space Telescope to monitor 
the UV variability of a sample of 17 galaxies with LINER nuclei and compact 
 nuclear UV sources. 
Fifteen of the 17 galaxies were observed more than once,
with two to five epochs per galaxy, spanning up to a year. 
We detect significant variability in most of the sample, with peak-to-peak
amplitudes from a few percent to 50\%. In most cases, correlated variations
are seen in two independent bands (F250W and F330W). 
Comparison 
to previous UV measurements indicates, for many objects, long-term variations 
by factors
of a few over decade timescales.
Variability is detected in LINERs with and without detected 
compact radio cores, in LINERs
that have broad H$\alpha$ wings detected in their optical spectra 
(``LINER 1's''), and in those that do not (``LINER 2s''). This
variability  demonstrates the 
existence of a non-stellar component in the UV continuum of all types, 
and sets a lower limit to the luminosity of this component. 
Interestingly, all the LINERs that have detected radio cores have 
variable UV nuclei, as one would expect from {\it bona fide}
AGNs.
 We note a trend
in the UV color (F250W/F330W) with spectral type -- LINER 1s 
tend to be bluer than LINER~2s.
This trend may indicate a link 
between the shape of the nonstellar continuum and the presence or 
the visibility of a broad-line region. In one target, 
the post-starburst galaxy NGC 4736,  
 we detect
variability in a previously noted UV source that is offset 
by $2\farcs 5$ ($\sim 60$~pc in projection) 
from the nucleus. This may be the nearest
example of a binary active nucleus, and 
of the process leading to black hole merging.          
\end{abstract}

\section{Introduction}

 Low-ionization
nuclear emission-line regions (LINERs) are detected in the
nuclei of a large fraction of bright nearby galaxies
(Ho, Filippenko, \& Sargent 1997a; Kauffmann et al. 2003).
Since their definition as a class by Heckman (1980), they have elicited debate
as to their nature and relation, if any, to active galactic nuclei (AGNs). 
Although 
the luminosities of most LINERs
are unimpressive compared to ``classical'' AGNs, a variety of observables
point to similarities and continuities between AGNs
and at least some LINERs. To list some of these, at least 10\% of LINERs
show weak, broad, Seyfert-1-like H$\alpha$ wings in their spectra 
(Ho et al. 1997b),
and Keck spectropolarimetry of several 
objects has revealed ``hidden BLRs'' (Barth, Filippenko, \& Moran 1999a,b),
similar to those seen in some Seyfert 2s (Antonucci \& Miller 1985; Tran 1995).
{\it Hubble Space Telescope} (HST) 
imaging shows that some 25\% of LINERs have compact, 
often unresolved 
(i.e., $\lesssim$ few pc), bright UV sources in their nuclei 
(Maoz et al. 1995; Barth et al. 1998).
Optical imaging with WFPC2 on HST of 14 LINERs (Pogge et al. 2000) suggests
that all nearby LINERs 
(including the 75\% that are ``UV-dark'', i.e., those that
do not reveal a nuclear UV source at HST sensitivity, $\sim
10^{-17}~\ergcmsA$) 
likely have such a nuclear UV source, but that it is
often obscured by circumnuclear dust.

In the radio,
at Very Large Array (VLA) resolution ($0\farcs 1\approx $ tens of pc), 
about half of LINERs display
unresolved radio cores at 2~cm and 3.6~cm
 (Nagar et al. 2000, 2002). At 6~cm, with 
Very Long Baseline Interferometer
(VLBI) resolution ($\sim 1$ pc),
these cores remain unresolved, strongly arguing for the presence of an AGN
(Falcke \& Biermann 1999; Falcke et al. 2000). 
The radio core fluxes have been found
to be variable by factors of up to a few 
in about half of the $\sim 10$ LINERs observed multiple times
over 3 years
(Nagar et al. 2002). A radio survey for 1.3~cm water megamaser
emission, an indicator of dense circumnuclear molecular gas, detected
LINER nuclei at the same rate as type-2 Seyfert nuclei (Braatz et
al. 1997). Such megamaser emission is seen only in AGNs.
Some LINERs have indications of a Seyfert-like ionization cone oriented 
along their radio axis (Pogge et al. 2000). 

At X-ray energies, Rosat HRI images showed compact ($<5''$) soft X-ray
emission in 70\% of LINERs and Seyferts (Roberts \& Warwick 2000) which, 
when observed with ASCA, were found to have a nonthermal
2-10 kev spectrum (e.g., Terashima, Ho, \& Ptak 2000). Arcsecond-resolution
Chandra observations by Terashima \& Wilson (2003) of 11 LINERs, each of which 
was
preselected to have a radio core, revealed an X-ray nucleus in all but
one case, and the nuclei  were generally (but not always) unresolved.

The super-massive black holes in the nuclei of most normal galaxies (e.g.,
Tremaine et al. 2002; Ferrarese \& Merritt 2000), 
many of which are also LINERs,
could be the remnants of ancient quasars/Seyferts, now
accreting at a low rate and/or radiating inefficiently (e.g., Reynolds et al.
1996), and producing these multiwavelength signatures.
If LINERs represent the low-luminosity end of the AGN phenomenon,
then they are the nearest and most common examples, and their 
study is germane to understanding AGN demographics and evolution,
and the X-ray background.

However, an unambiguous optical/UV link between the LINER and AGN 
classes has remained elusive. Maoz et al. (1998) analyzed the HST UV spectra
of seven ``UV-bright'' LINERs and showed that,
in at least some of them, most or all of the compact UV continuum
 emission is produced
by a cluster of massive stars,
 whose energy output may be  sufficient to account for the
optical recombination lines. Even in the few objects showing broad,
quasar-like emission lines, one cannot say conclusively 
whether the UV continuum source is stellar or nonstellar, because the
broad emission lines coincide in wavelength with the main expected stellar 
absorptions.
 X-ray and radio data have provided convincing evidence for 
the presence of AGNs in {\it some} of these objects. 
However, they cannot identify the source of
the optical emission lines, which are excited by UV radiation 
(beyond the Lyman limit), and
the entire LINER definition
rests on the ratios of these emission lines. There is thus the
possibility that the LINER phenomenon 
and central black holes are not physically
connected, but simply coexist in many
galaxies because both are common.

Variability is one of the defining properties
of AGNs. 
Variability
can reveal an AGN origin of an emission component, 
even when broad lines are not detected
and much of the continuum emission is produced by a nuclear star cluster.
For example, in both of the LINERs
 NGC~4569 and NGC~404, Maoz et al. (1998) showed that
the UV spectrum has the broad absorptions  in C~IV~$\lambda1549$ 
and N~V~$\lambda 1240$ due
 to winds from O-type stars. However,
the relative shallowness of these features in NGC~404 
means the O-star light is
diluted by another component. This component could be lower-mass stars
(B and A dwarfs) in the same cluster, or it could be a featureless
AGN component. 
Repeated observations could potentially detect
UV variations, thereby exposing an AGN component in the UV,
even when much of the continuum emission is produced by a nuclear star cluster.

Indeed, there have been several reports of UV variability in LINERs.
Barth et al. (1996) compared their HST/FOS spectrum of NGC~4579 to the
HST/FOC F220W measurement of Maoz et al. (1995), and found a factor of 3
decrease in the 
flux of the central source over the 19-month period between the observations.
Cappellari et al. (1999) compared FOC observations of NGC~4552 taken
in 1991, 1993, and 1996, and found a factor of 4.5 brightening between
the first two epochs, followed by factor of 2 dimming between the last two
epochs. 
While these detections of UV variability were suggestive, they were
not conclusive from the technical aspect, because the observational
setup was different at each epoch. In the case of NGC~4579, one is
comparing a broad-band measurement to a spectroscopic one, where
aperture misplacement is always a danger. In the case of NGC~4552,
the three measurements were both pre- and post-HST-spherical-abberration
correction,
 and in different FOC
formats, having different dynamic and non-linearity ranges, and different 
fields of view.
Other indications of UV variability in LINERs have been indirect,
 e.g., the appearance of broad (sometimes double-peaked)
Balmer emission-line wings in some galaxies with LINER spectra
(e.g., Storchi-Bergmann et al. 1995).

If the reported UV variations in LINERs are real, it would mean that:\\
1. Some of the UV emission of these LINERs is of an AGN nature;\\
2. LINER variations are common (since UV variations were detected
in the few galaxies that were observed more than once); and\\
3. Large-amplitude (factor $\sim 3$) variations on few-year 
timescales are the norm.
This would contrast with Seyferts 1s, where typical UV variations 
are $\lesssim 2$, and quasars, where variations of only tenths of a magnitude
are most common (e.g., Giveon et al. 1999).\\
Confirming (or refuting) the above results on a carefully-selected
sample could therefore provide the missing link between LINER emission
 and AGNs, and supply important new input to the phenomenology
of AGN variability and its dependence on luminosity.

Quantifying the stellar and AGN contributions to the UV is also
important for correctly
comparing  the continuum emission of low-luminosity AGNs with models.
While in quasars and Seyferts the UV continuum in generally attributed
to a standard thin accretion disk (e.g., Shakura \& Sunyaev 1973),
in low-luminosity AGNs the nature
of the UV continuum is still a matter of debate.
In models
such as advection/convection dominated accretion flows (ADAFs/CDAFs; 
e.g., Quataert et al. 1999), the emission is
from the hot accretion flow itself. Alternatively, Yuan et al. (2002)
and Falcke et al. (2004) have postulated that not only the radio
emission,
but also the optical/UV/X-rays in low-luminosity objects could be nonthermal
emission, likely from the jet itself, with the jet emission becoming 
dominant as the disk becomes radiatively
inefficient. In the latter picture, one postulates a transition from a
thermally dominated spectrum to a non-thermal (possibly jet-dominated)
spectrum as one goes from  black holes accreting near the Eddington
limit (quasars and Seyferts) to black holes with sub-Eddington
accretion. Stronger UV variability in low-luminosity AGNs
could be a
signature of this transition.

In the current paper, we present results from 
a monitoring program using HST to search
 for UV variability in a sample of LINERs over timescales of weeks to 10 
years. 
Compared to previous, serendipitous,
detections of variability, our program was designed to study 
 this question systematically, using a stable 
observational setup 
and a representative sample
of UV-bright LINERs. 

\vskip 2cm

\section{Sample and Observations}

\subsection{Sample}
Our sample includes all objects
seen to have compact central UV sources in existing HST data
as known by us in 2001, 
 and classified
optically as LINERs by Ho et al. (1997a) based on their optical emission line
ratios.\footnote{Our sample inadvertently excluded
NGC~4303, a UV-bright nucleus (Colina et al. 1997,
2002), classified by Ho et al. (1997a) as an H~II nucleus, but 
which higher spatial resolution spectroscopy by Colina \& Arribas
(1999) identified as a LINER}
The 17 LINERs in the sample
include a variety of LINER sub-types: LINERs having broad 
H$\alpha$ wings (which we will designate ``LINER 1s''), 
and those having only narrow emission 
lines (``LINER 2s'');
LINERs whose UV spectra show signatures of massive stars, and those that
do not; and some LINERs that, in terms of optical classification, are
borderline with Seyferts or with H~II nuclei. We will refer to all
these objects collectively as either LINER~1s or LINER~2s, depending
on the presence or absence of broad 
H$\alpha$ wings.\footnote{Ho et
  al. (1997a) designated LINERs with broad H$\alpha$ wings as
  LINER~1.9 objects. Since this is the only kind of type-1 LINER in the
  sample of Ho et al. (i.e., there are no known examples of
  LINER~1.2, 1.5, etc.), we will simply refer to LINER~1.9s as
  LINER~1s. Three of the objects in our sample are classified by Ho et
al. as Seyferts, since their narrow
emission line ratios [OIII]~$\lambda 5007$/H$\beta$ are 
above the defining border between LINERs and Seyferts by $\sim 30-40\%$
(for M81 and NGC~3486) and by a factor  $\sim 3$ (for NGC~4258).
The border is somewhat arbitrary (see, e.g., the distribution of
emission-line nuclei from the Sloan Digital Sky Survey 
on the diagnostic diagrams shown by Kauffmann et al. 2003), so 
we consider the former two objects
also as borderline LINER/Seyfert cases, and the latter as a
low-luminosity Seyfert.}
  VLA imaging at 2~cm
and 3.6~cm has revealed
a radio core in 11 of the objects, and variability has been detected 
(Nagar et al. 2002) in five of those that have been
monitored.
Tables 1 and 2 list the objects in the sample and summarize some of 
their previously known properties. 

\subsection{Observations}
Imaging of the sample was carried out
with the HST Advanced Camera for Surveys
(ACS) with its High Resolution Camera (HRC) mode. The field of view of this
CCD-based instrument is about $29''\times 25''$, with a scale of 
$0\farcs 0284\times 0\farcs 0248$~pixel$^{-1}$. Each target was imaged in the
F250W band ($\lambda_{\rm central}\approx 2500$~\AA, FWHM$\approx 550$~\AA)
with exposure times ranging from 5 to 25~min, depending on target brightness, 
and in the
F330W band ($\lambda_{\rm central}\approx 3300$~\AA, FWHM$\approx 400$~\AA)
with an exposure time of 5~min. (The brightest target,  NGC~4569, was exposed
for just 1~min in each band.) The exposure time was split between two 
equal exposures that were used in the data reduction process to reject 
cosmic-ray events. Objects were repeatedly scheduled using HST's
Snapshot mode, i.e., these short exposures were chosen by the HST
schedulers, as dictated by convenience, in order to fill gaps left in the 
schedule after normal-mode observations had been scheduled. This means that
both the number of epochs at which a given target was actually observed and
the spacing between epochs were largely random. Between July 1, 2002 and 
July 2, 2003, 15 of the 17 LINERs were observed from
 two to five times each. Two objects,
NGC~404 and NGC~1052, were observed only once. HST failed to acquire
guide stars in two exposures, of
NGC~3642 on December 17, 2002, and of NGC~3486 on February 13, 2003. 
The resulting failed data will be ignored here. We supplemented our data
with archival data obtained with the same observational 
setup for two of the objects,
using the F330W filter: one epoch for NGC~3486 from June 3, 2003, 
and one for NGC~4258 from December 7, 2002.
Table 1 lists the exposure
 times and epochs of each target.

\begin{figure}
\epsscale{1.2}
\plotone{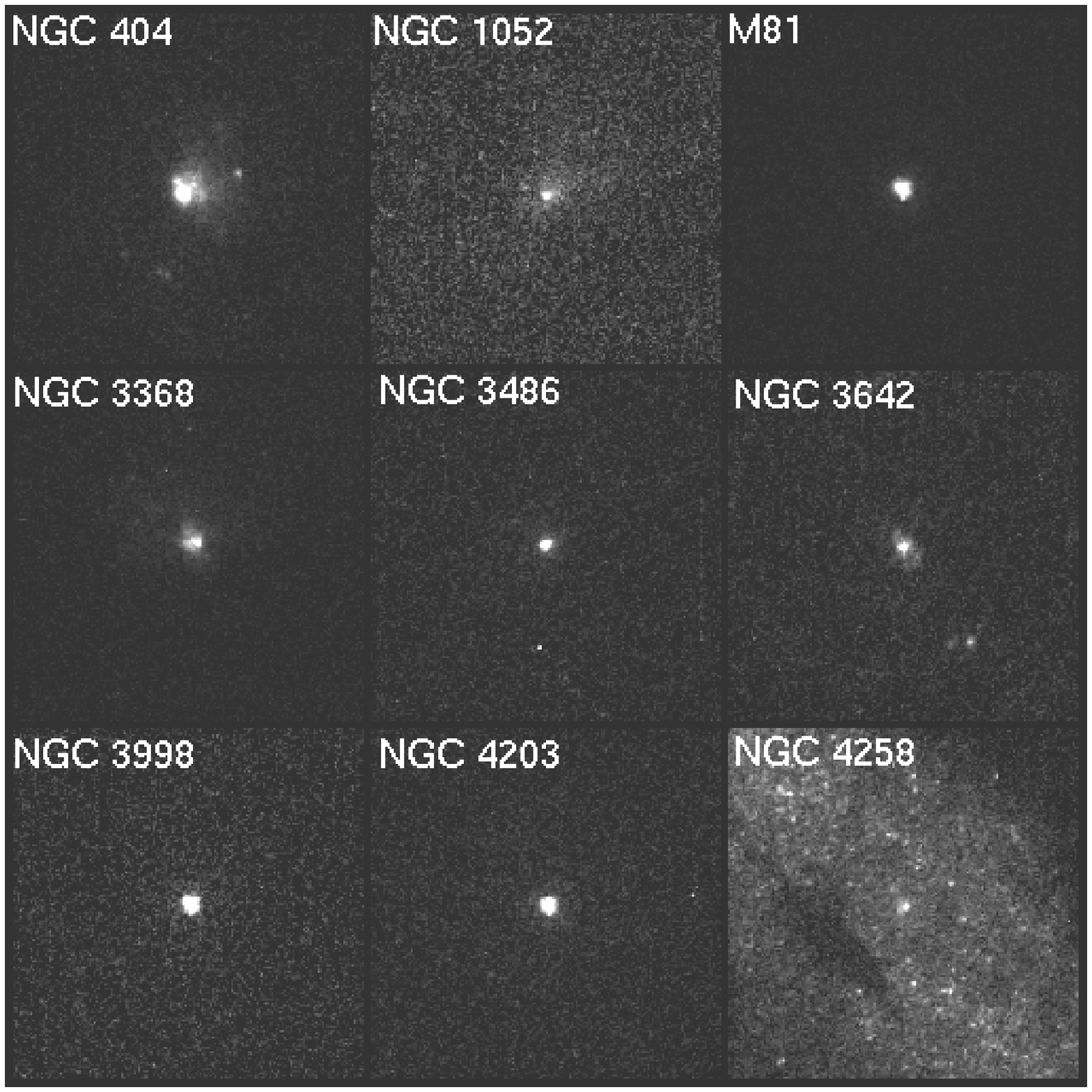}
\end{figure}
\begin{figure}
\label{imagemosaic}
\epsscale{1.2}
\plotone{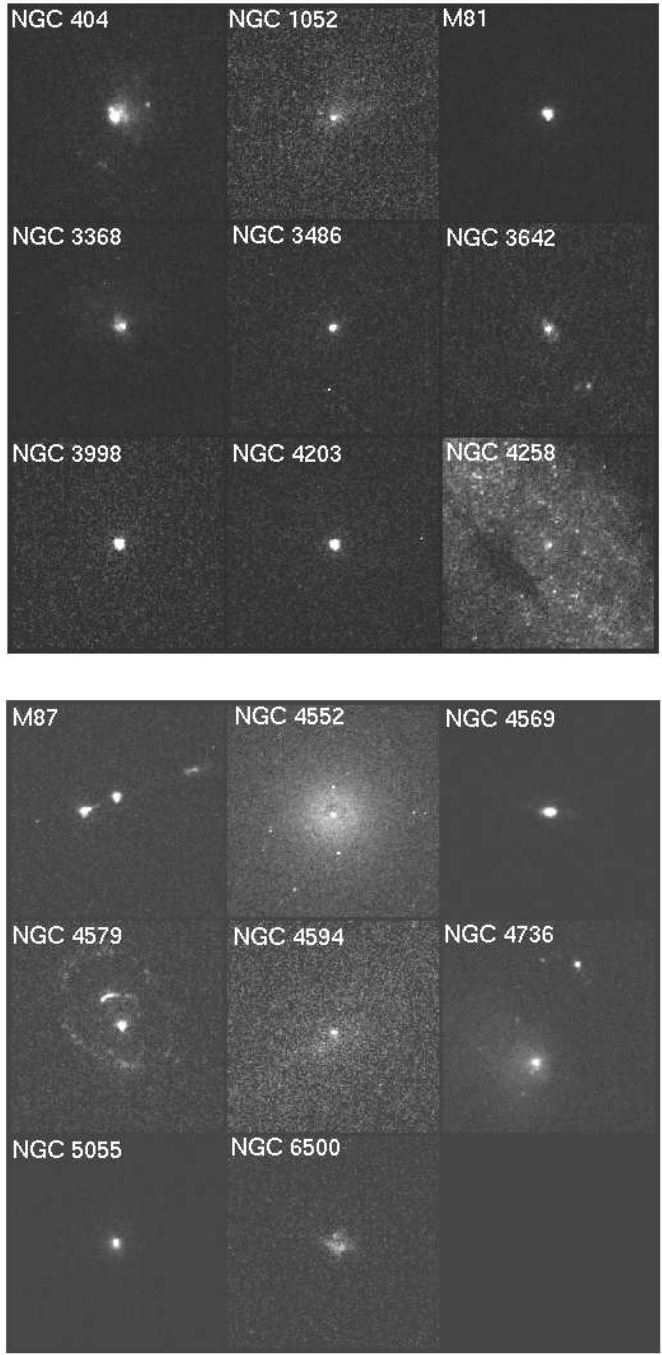}
\caption{Sections of the ACS/HRC F250W images, $5''$ on a side,
 of each of the galaxies in the sample. The nuclei are at the center of
 each image, except in the case of M87 and NGC~4736, in which the
 nuclei are
 slightly offset to the lower left
 so as to include in the image the jet and the
 off-nuclear UV source, respectively. Note that an
 unresolved object is the only significant UV source in the central
 regions of most of these bright, nearby, galaxies.}
\end{figure}

Figure 1 
 shows ACS image sections around each of the LINER nuclei. Compared
with previous UV images of these objects with the FOC, WFPC1, and WFPC2 
instruments, the improvement in resolution, sensitivity, and linearity 
reveals, in some cases,
 fine structures and details that were not clearly seen before.
 However, the morphologies characterized by isolated, compact or unresolved, 
nuclear UV sources
are consistent 
with the previous images by Maoz et al. (1995) and Barth et al. (1998).
This simple morphology also facilitates 
 the photometric measurements described below.   

\section{Analysis and Results}

\subsection{Data reduction}
Data were reduced automatically by the Space Telescope Science
 Institute (STScI)
 pipeline. This includes
bias and overscan subtraction, cosmic-ray rejection and combination
of the two split exposures, dark subtraction, flat-fielding, and geometric
distortion correction.
To assure uniform reduction using
 the latest available flats and distortion-correction algorithms,
the entire dataset was re-retrieved from the HST archive 
 and processed on-the-fly on  May 4, 2004.

\subsection{ACS UV photometric stability} 
Of great concern in a program such as this is the photometric stability of
the camera. This concern is heightened by the fact that, given the small field
of view, the short exposures, and the old stellar population 
of the galaxy bulges in which LINERS are preferentially found, the compact
nucleus is generally the only bright feature in the image, and 
relative photometry is not possible. 
Indeed, WFPC2, the ACS's predecessor CCD imager on HST, suffers
from severe and variable molecular contamination on its front window, which 
causes large variations with time in UV sensitivity. ACS was expected not to
be afflicted by such a problem, since it does not have a cold window on which
contaminants can condense. However, as our program was executed on the first 
observing cycle after the installation of ACS, the actual in-flight UV 
sensitivity stability was not known, and must be addressed when assessing the
 reality of any detected UV variations.

Fortunately,  STScI 
 staff carried out a program to monitor the UV photometric 
stability of ACS/HRC, including the two filters we used, contemporaneously 
with our program. As reported by Boffi, Bohlin, \& de Marchi (2004), 
the open star cluster NGC~6681
was observed 19 times from May 2002 to July 2003. Observations were
roughly bi-weekly until mid-November 2002, then paused, and resumed
in mid-February 2003, roughly once a month. Not only was the 
observational setup identical to ours, but the total exposure time 
per filter (140~s split into two exposures) 
was comparable to the one we used (300~s for most targets), and the 
brightnesses
of the stars (20,000-30,000 total counts per star within an 8.5-pixel radius) 
in this test field were similar to those of the compact nuclei in our program.
Boffi et al. 
show light curves for eight of the stars in the field, and report that
the UV sensitivities in both F250W and F330W are stable ``to 1\%''. 

To obtain a more quantitative estimate of the photometric stability,
we used the plots of Boffi et al. (2004)
 to calculate the actual rms scatter of each star's
light curve in each filter. We also measured the brightness of each of these
stars in several of the reduced images of NGC~6681 that we retrieved from
the HST archive. We then subtracted, in quadrature from the rms scatter
of each light curve, the readout noise due to the 
pixels within an 8.5-pixel radius in two exposures, and the Poisson noise
due to the total counts, to obtain the remaining scatter due to other
sources of noise. We designate this remaining noise as
the ``photometric scatter''.
We find that, among  the 16 stellar light curves (8 stars
in two bands each), the photometric scatter ranges from zero (i.e., the 
rms of a light curve is at the level expected from Poisson noise and
readout noise alone) up to 1.1\%. Four of the light curves have photometric
scatter near zero, eight have scatter
from 0.4\% to 0.7\%, and four have scatter of about 1\%. 
We do not find a correspondence between the photometric
scatter of a light curve and any obvious parameter, such as the 
identity of a star, its
position on the chip, its brightness, or its color. In fact, some of the stars
with the highest scatter in one band have the lowest scatter in the other
 band, even though at each epoch the two bands were obtained consecutively,
with shifts of only a few pixels between exposures. Since a sizeable
fraction (25\%) of the stellar light curves show photometric scatter
of about 1\%, and since the first purpose of
the present study is to test the null hypothesis that  LINERs do not vary, 
we will conservatively assume that the ACS/HRC has a photometric rms
scatter of 1\% in both the F250W and the F330W bands.
We will adopt this figure as the photometric calibration
uncertainty of our measurements, to be combined with the other, statistical,
 sources of error. 

\subsection{Photometry}
We used IRAF\footnote{IRAF (Image Reduction and Analysis
Facility) is distributed by the National Optical Astronomy Observatories,
which are operated by AURA, Inc., under cooperative agreement with the
National Science Foundation.}
 to perform aperture photometry of the nuclear source in 
each image. Counts were summed within a
 10-pixel-radius ($0.27''$) 
aperture centered on the source. The background level was
determined from the median counts in an annulus at radii of 14 to 18 
pixels. Experimenting with the more constant among the nuclear sources
(e.g., NGC~4569),
we found that an aperture radius of $>8$~pixels is required in order
to obtain photometric stability of better than 1\% between epochs. This
is consistent with the use of 8.5-pixel-radius apertures by Boffi et al. (2004)
in the photometric stability tests described above.  
Errors were calculated by combining in quadrature the Poisson errors of
the counts, the readout-noise errors from the pixels within the aperture
in the two split exposures (assuming a readout noise of 4.71~e~pixel$^{-1}$),
and the adopted photometric scatter of 1\%. Counts and their errors were
multiplied by 1.25 (for F250W) and by 1.18 (for F330W) to
correct for the finite aperture radii, based on the point-source encircled
energy curves in the ACS Data Handbook (Pavlovsky et al. 2004).
Finally, count rates were 
converted to flux densities using the conversion given by the 
PHOTFLAM keyword in the image headers,
1~e~s$^{-1}=4.781\times 10^{-18}~\ergcmsA$ (F250W), and    
1~e~s$^{-1}=2.237\times 10^{-18}~\ergcmsA$ (F330W). This conversion
assumes a spectral shape that is flat in $f_{\lambda}$, which is a
reasonable approximation for these objects -- their UV colors (see
below) imply a spectral slope in the range $-0.4< \alpha < 0.4$, for
an assumed spectral shape $f_{\lambda}\propto \lambda^{\alpha}$.   

Three galaxies, NGC~4736, NGC~5055, and NGC~6500,
 merit separate mention. NGC~4736, apart from its nuclear UV source
(which is clearly centered on a diffuse, centrally peaked stellar light
 distribution) 
displays a second UV point source, $2\farcs 5$ north of the nuclear UV
 source. The flux from the off-nuclear
source, which we designate NGC~4736b, was measured exactly as for the other
nuclear sources. The off-nuclear source will be further discussed below.
 NGC~5055, has a central source which is resolved, with an
observed full-width-half-maximum (FWHM) of about 5.5~pixels ($0.15''$),
as opposed to the 2-3~pixel FWHM typical of point sources. 
In its second epoch, on March 12, 2003, it is significantly {\it more}
 extended in both bands, 
with a FWHM of 7~pixels, perhaps due to spacecraft jitter.
To prevent both the normal large width and the anomalous  epoch 
from adversely affecting the photometry, we used a 13-pixel aperture
in this case, which increases the flux by 20\% but eliminates a 
spurious 3\% decline at this epoch.  
NGC~6500 does not have a clearly defined nuclear source.
Instead, it has a diffuse central light distribution, on which are superposed
a number of faint sources, some compact and some extended. Although this 
structure was already known from previous imaging with WFPC2 by Barth et al.
(1998), we included this galaxy in the sample since it has various known
AGN properties (a radio core -- Nagar et al. 2000; 
a possibly nonstellar UV spectrum --
Barth et al. 1997), keeping
in mind the possibility that one of the faint sources in the WFPC2 image
could have been the active nucleus, perhaps temporarily in a low state.
To encompass within the aperture the diffuse nuclear light from this
galaxy, we used an aperture radius of 20, rather than 10, pixels.

\subsection{Light curves}

\begin{figure}
\epsscale{2.6}
\plottwo{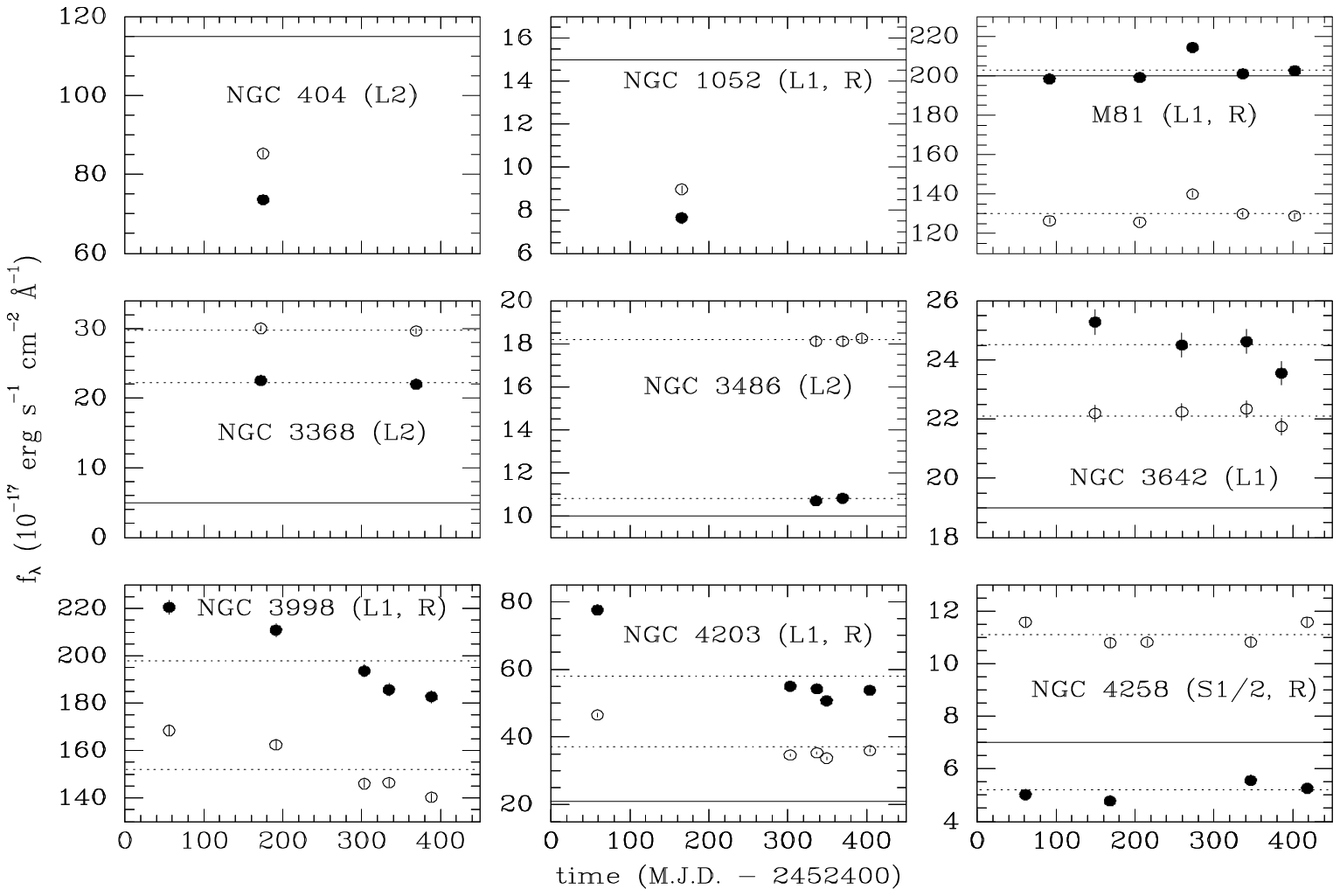}{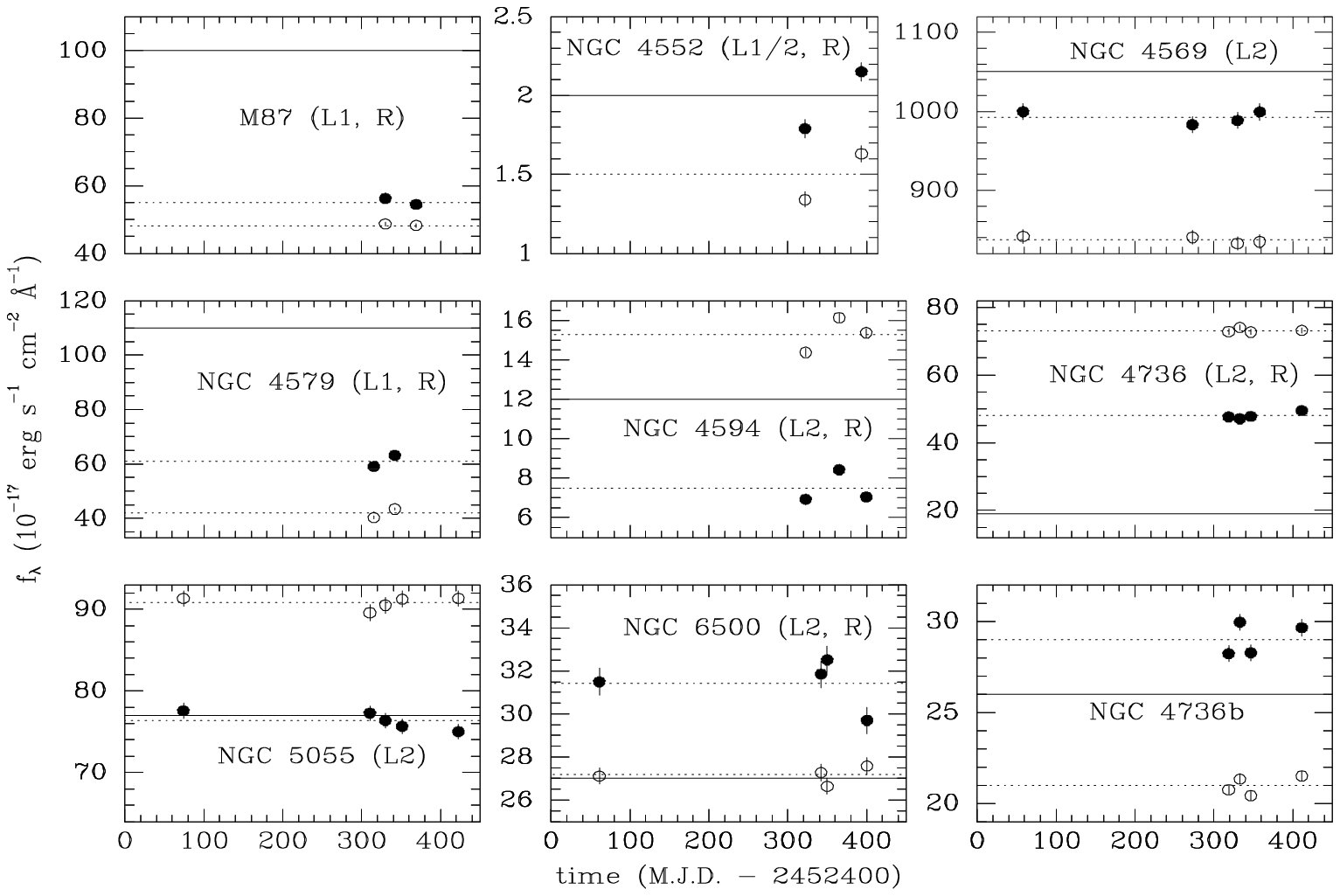}
\label{lightcurves}
\epsscale{1.3}
\caption{UV light curves in F250W (filled circles)
and F330W (empty circles) for each of the 17 nuclei in the sample, plus
the off-nuclear source NGC~4736b. The period shown corresponds to May 7,
2002 through July 30, 2003. Objects are labeled in parentheses as
L1 for type-1 objects (including transition LINER/Seyfert objects),
L2 for  type-2 objects (including transition LINER/Seyfert objects
and transition LINER/H~II objects), and R for objects with a detected
radio core. The Seyfert nucleus
NGC~4258 and the LINER NGC~4552, have broad
components in both the permitted and the forbidden lines, and 
are labeled S1/2 and L1/2, respectively. Dotted lines show the time-averaged
mean flux in each band, and solid lines show one of the 
``historical'' (1991-1997) flux levels measured previously  
for these objects at bandpasses similar to the F250W band, 
usually at 2200~\AA~. See \S4, Notes on Individual Objects, for details.
A historical level is not shown for NGC~3998, as it is $\sim 5$ times
higher than the 2003 level.
Many of the objects display significant short-term ($\lesssim 1$~yr) 
variations, correlated between both UV bands, and large-amplitude 
long-term variations.}
\end{figure}

 The fluxes at every epoch are included in Table~1. 
Figure~2 shows the light curves in F250W (filled circles)
and F330W (empty circles) for each object. Every object is designated
as L1, for type-1 objects (including transition LINER/Seyfert objects), or
L2, for  type-2 objects (including transition LINER/Seyfert objects
and transition LINER/H~II objects),
 and is marked with an ``R'' if it has
a detected compact flat-spectrum radio core. The Seyfert nucleus
NGC~4258 and the LINER NGC~4552, unusual objects that have broad
components in both the permitted and the forbidden transitions
(see below, and in Notes on Individual Objects), 
are labeled S1/2 and L1/2, respectively.
The horizontal 
dotted lines are plotted at the time-averaged
mean flux value in each band. The solid lines show one of the 
flux levels measured previously (1993-2000) for these objects at 
bandpasses similar to the F250W band, usually at 2200~\AA. 
While straightforward comparison
to the currently measured levels is difficult (see \S 1), very large
variations, of a factor of a few, between these ``historical'' measurements
and the current ones are probably real. Such a comparison is discussed 
in each case in \S4, ``Notes on Individual Objects''.
 
In Table~2 we list 
some statistics derived from the light curve of each object. These
include: the number of epochs; the time-averaged mean flux in each band; 
the $\chi^2$ per degree
of freedom of the data in each band, relative to a model with a
constant (non-variable) flux at the mean level -- this number appears in 
boldface for the cases that are variable at $>95\%$ confidence; 
the peak-to-peak variation, with the typical error subtracted in quadrature 
 (if negative, the peak-to-peak variation is set to zero); 
the  time-averaged mean UV color 
$f_{\lambda}$(F250W)/$f_{\lambda}$(F330W), after correction for Galactic
reddening, assuming the $B$-band extinction values of Schlegel et al. (1998)
and the Galactic extinction curve of Cardelli et al. (1995) with
the parameter $R_V=3.1$; the color change between  
the two epochs with extreme fluxes -- 
[$f_{\max}$(F250W)/$f_{max}$(F330W)]/
[$f_{min}$(F250W)/$f_{min}$(F330W)]; and its uncertainty. 

\subsection{Distances}
We have compiled from the literature recent distance estimates to all
the galaxies in the sample. In 11/17 cases, the distances are
based on ``modern'' methods -- Cepheids, surface-brightness
fluctuations, tip of the red giant branch, Tully Fisher, and maser
proper motion. Several of the galaxies have distances from several
different 
methods, which always agree to better than 10\%, in which cases we
have used the averages. Our adopted distances are listed in Table~2,
along with the literature sources on which they are based. 
We have used these distances, and
Galactic extinction corrections as described above, to compute
monochromatic luminosities  at 2500~\AA, which are also given in Table~2.

\subsection{UV Variability}
Inspection of  Figure~2 and Table 2 reveals a number of new
results. First, in the F250W band, among the 16 objects with multiple
epochs (the 15 galaxies with multiple epochs, including the double nucleus
in NGC~4736), significant variability  at greater than
the 95\% confidence level, based on $\chi^2$, is detected  
in all but four cases: NGC~3368, NGC~3486, NGC~4569, and NGC~5055 
(plus M87, that  varied at 94\% confidence.
The apparently significant variations in NGC~6500 are uncertain -- see below).
In the F330W band, eight objects reveal significant changes, and
all of these vary in F250W as well. 
Whenever
significant variations are detected in both bands (in eight objects), 
the variations are  
correlated. Significant variations range in peak-to-peak amplitude
(expressed as a fraction of the mean flux) from 3\% to 46\%, with a
median of 7\%, in F250W, and from 5\% to 34\%, with a median of 11\%, in
F330W. (Note that these statistics are affected by the different
number of epochs and time intervals for each object.)

As summarized in \S1, UV variability in some of these LINERs has
been reported before. However, this is the first time such variability is seen
on relatively short timescales, and it is detected using an unchanging, 
and photometrically very stable, observational setup (\S 3.2). 
The variable flux provides
a firm lower limit on the AGN contribution to the UV flux at each band.
We see that variability, and hence an AGN contribution,
 exists in some members of both LINER classes, 1 and 2. This situation is
distinct from the one in Seyfert galaxies. In Seyferts, the AGN 
continuum that is
visible in type-1 objects is
obscured in type 2s, in which the observed UV continuum is sometimes
scattered AGN light and sometimes produced 
by young stars in the circumnuclear region (e.g., Gonz\'alez-Delgado
 et al. 1998), and, in either case, is not expected to be variable on 
$\approx 1$~year or shorter timescales. Our finding that variability is seen
 in at least some LINER 2s suggests that the ``unified scheme'', believed 
to apply to Seyferts, may not always apply to LINERs. 
    
In terms of longer timescale variability, comparison of the 
UV flux levels we measure 
to the ``historical'' ones shown with a solid line in 
Figure~2 reveals
likely large-amplitude variations even in some objects that were not seen to 
vary during the present campaign, either because they were sampled too
closely or too infrequently, or because they were temporarily inactive. 
These include
NGC~404, NGC~1052, NGC~3368,  and M87 (see \S4, for
details). Indeed, the nucleus of M87 was recently monitored by Perlman et al. 
(2003) using the ACS/HRC with the F220W filter, and was shown to vary  
in flux, consistent with previous reports of optical variability 
by Tsvetanov et al. (1998). 
 This leaves only three out of the 17 LINER nuclei that 
appear to be constant on both short (months) and long (up to 10 years)
timescales. These constant objects are NGC~3486, NGC~4569, and NGC~5055.
 We can conclude, therefore, that UV variability is a common
property of the majority of LINERs.  

The nuclear UV source in NGC~4736 appears variable in F250W, but only at the 
95\% confidence level, and is constant in F330W. 
However, a surprising result is that the off-nuclear 
source, NGC~4736b, is clearly variable in both bands. This raises the 
possibility that the off-nuclear source is the active nucleus of a second 
galaxy in the final stages of a merger with NGC~4736, or perhaps
it is related to jet activity originating in the nucleus. 
This is discussed in more detail below.  

\begin{figure}
\label{colorhist}
\epsscale{1.2}
\plotone{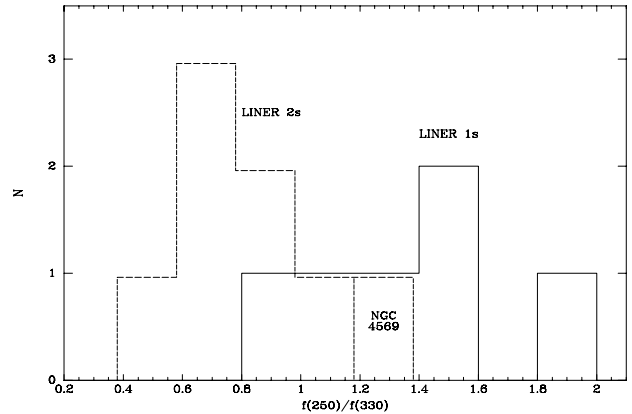}
\caption{Histograms of Galactic-reddening-corrected 
UV color, $f_{\lambda}$(F250W)/$f_{\lambda}$(F330W), for type-1 objects
(solid line) and type 2 objects (dashed line). Type-1 objects appear to
be generally bluer than type-2's. NGC~4569, the bluest
type-2 object (see text), is labeled.}
\end{figure}

\subsection{UV color of type-1 and type-2 LINERs}
Inspection of Table 2 reveals another interesting result. In terms of
the Galactic-reddening-corrected 
UV color $f_{\lambda}$(F250W)/$f_{\lambda}$(F330W),
there is an apparent trend that LINER 1s 
are, on average, blue, and LINER~2s are red.
This is illustrated in Fig.~\ref{colorhist}, showing 
histograms of the UV color for LINER~1s and LINER~2s.
The populations (at least as probed by our small sample) seem to
overlap at $f_{\lambda}$(F250W)/$f_{\lambda}$(F330W)$\approx 1$.
 We have excluded
four objects from this plot:\\
{\bf  NGC~4258}, which we have listed as an object of uncertain type,
has the reddest nucleus in our sample. 
Pogge et al. (2000) have presented
evidence that this object is a borderline case between UV-bright and UV-dark
objects, and is likely partially obscured and reddened by foreground
dust. Furthermore, its narrow [OIII]/H$\beta$ ratio of 10 puts it
firmly in the Seyfert regime (Ho et al. 1997a), distinct from the
other nuclei in the sample, which are LINERs or borderline LINERs.
Its broad H$\alpha$ wings, which could give it a type-1
classification, are also peculiar in the sense that both the permitted and the 
forbidden lines in its spectrum 
have broad bases, especially in polarized light (Barth et al. 
1999b).\\
{\bf NGC~4552} was classified by Ho et al. (1997a) as ``T2:'',
meaning a transition object between a LINER and an H~II nucleus, with
no evidence of a broad H$\alpha$ component, and an uncertain classification.
The uncertainty is driven by the weakness of the emission lines, which
in the ground-based optical spectrum are superposed 
on bright stellar emission from the center of the galaxy.
Since the narrow emission lines are so faint, an even-weaker broad component
would have been impossible to detect, and  the classification 
of NGC~4552 as a type-1 or type-2 LINER was limited by the signal-to-noise
ratio of the spectral data (L.C. Ho, private communication).
Indeed, Cappellari et al. (1999) analyzed
an HST/FOS spectrum of NGC~4552, in which, at HST resolution,
 the nucleus can be
better isolated from the surrounding starlight. They found that the
emission lines have a significant broad component, with velocities of 
3000~km~s$^{-1}$, typical of type-1 objects. However, this broad
component was present in both the permitted {\it and} the
forbidden lines, in contrast to other type-1 AGNs, but reminiscent of
the situation in NGC~4258. It is therefore unclear whether NGC~4552
should be considered a type-1 or a type-2, and we exclude it from the
UV-color analysis.
[Incidentally, Cappellari
et al. (1999) classified NGC~4552, using the narrow-line ratios
measured with HST,  as a borderline case between a LINER 
and a Seyfert, rather than as a
 transition case between a LINER and an H~II nucleus.] 
We note that a problem in detecting a faint broad component 
does not exist at  the same level 
for the other LINER~2s in our sample, which are 
considerably brighter than NGC~4552.
 In the UV as well, NGC~4552 is, by far, the 
faintest source in our sample, and one of the
least luminous.\\
{\bf NGC~4736b}, the off-nuclear source, 
 was excluded from the UV-color analysis 
because we know nothing about its nature, 
let alone if it is a type-1 or type-2 object.\\
{\bf  NGC~6500} was excluded from the plot because it has no obvious
nuclear source whose color can be measured.\\
This leaves 14 objects in the plot -- six LINER~1s and eight LINER~2s.    

One LINER~2, NGC~4569, is very blue, contrary to the ``1=blue/2=red''
trend. However, it is a peculiar object in other respects as well.
Maoz et al. (1998) have shown that
its UV spectrum is completely 
dominated by the light from O-type stars, and it has
failed to show AGN characteristics in any spectral band. In the present
study, it is one of the three objects that are not variable even on
decade timescales. Barth \& Shields (2000) have calculated photoionization
models specifically for this object, and have shown that, under particular
conditions, a young stellar cluster can produce the observed optical
emission-line spectrum. (Specifically, a sufficient number of Wolf-Rayet
stars must be present, implying both an instantaneous starburst and 
a current age of 3-5~Myr.) Perhaps NGC~4569 is a peculiar case of 
a starburst-driven LINER, and therefore differs from other LINER~2s 
also in its UV color.  

If our apologies for the excluded and the outlier nuclei are justified,
and there is a true UV color distinction between LINER~1s and LINER~2s,  
this may be the first independent observable that can predict the existence of
broad H$\alpha$ in a LINER, even if only for LINERs that are at the
extreme red and blue ends of the color distribution. It is tempting to speculate that the
redness of LINER 2s is produced by intervening dust, and that this
dust somehow obscures the broad-line region, in some analogy to the
unification schemes applicable to Seyfert 1s and 2s. However, 
such a scenario, at least in its simplest version,
 will not work.  
There is, at most, a factor $\sim 2$ difference in 
the UV color ratio of the LINER 1s and LINER 2s (see Fig. \ref{colorhist}).
Assuming a dust screen with a standard Galactic extinction curve,
a ratio of 2 between the continuum attenuations at 2500~\AA~ and 3300~\AA~  
implies an attenuation of flux in the H$\alpha$ spectral
region by only a factor of 2.4. If anything, one would expect the 
broad emission lines, which come from an extended region, to be 
even less attenuated than the continuum. In other words, in LINER~2s
the ``hidden''
broad H$\alpha$ flux  would be lowered by a factor $\sim
2$, relative to LINER 1's. Such a small reduction
would probably not hinder the detection
of broad components in type-2 objects, many of which are quite bright.
To further investigate a possible role of dust, we have searched for
correlations between the UV color we
measure and the Balmer decrements of the objects in the
sample, or the estimated host galaxy extinctions, 
as tabulated by Ho et al. (1997a). No trend was found. 
There is also no obvious relation between 
UV color and radio power, as tabulated by Nagar et al. (2002).
Thus, the UV color may indicate some other link 
between the shape of the nonstellar continuum and the presence or 
the visibility of a broad-line region.

 On the other hand, even in
high-luminosity AGN, there is seldom good agreement between dust
extinction measures at different wavelengths, probably because of a
combination of optical depth and geometry effects. Furthermore,
considering the challenge of detecting weak broad H$\alpha$ wings, a
factor 2 reduction may play some role after all. For example, by
isolating galactic nuclei from the surrounding stellar light using the small
spectroscopic apertures possible with HST, several LINERs, although previously
classified as LINER 1s from the ground, have revealed also broad 
double-peaked H$\alpha$ profiles (Ho et al. 2000; Shields et al. 2000).  
Thus, further work is required both to confirm the suggested trend of 
UV color with type, and to understand the effect. 

\subsection{UV color variation}
In terms of the temporal variation in UV color of the sources, 
one can see from Table~2 that the color remains constant, to within 
the errors, when the flux varies. The sole exception is NGC~4203, which
is significantly bluer when it is brighter, as is commonly observed
in many high-luminosity AGNs (e.g., Giveon et al. 1999). Note that this
is also the object with the largest amplitude of variations. Thus,
it may be that similar color changes occur in some of the other
 objects, but the
color changes are too small to be seen when the flux variation
amplitude is small. Indeed, Tsvetanov et al. (1998) have reported,
for M87 at optical wavelengths, a continuum that bluens as it brightens.
However, two of the galaxies with relatively large variation amplitudes,
NGC~3998, NGC~4579, keep their color constant to within a few percent
 when they vary, so the brighter-bluer phenomenon is not universal. 
     
\section{Notes on Individual Objects}

\noindent{\bf NGC~404} -- HST UV spectroscopy of this LINER~2 nucleus by 
Maoz et al. (1998) showed clear absorption signatures of OB stars.
However, the relative shallowness of the absorptions meant that the 
light from massive stars was diluted by another component, comparable in flux, 
which could be
a featureless AGN continuum, or the light from less massive stars in 
an aging or continuous starburst. Nagar et al. (2000) did not detect 
a radio core in this galaxy, to a limit of 1~mJy. 
A {\it Chandra} X-ray image presented
by Eracleous et al. (2002) shows a compact nuclear source
with 0.5-8 keV luminosity of  $1\times10^{37}$~erg~s$^{-1}$,
surrounded by some faint blobs.

In our new ACS images, the nucleus consists 
of a compact core on top of a  diffuse halo of
$\sim 0.5''$ diameter, and several surrounding faint sources. 
Since only one epoch was obtained for this object,
we cannot say anything about short-term variability. However, the 
2500~\AA~ flux is $\approx 60$\% 
of the level measured by the 1994 spectroscopy 
analyzed by Maoz et al. (1998; $115\times 10^{-17}~\ergcmsA$). 
This difference is likely real, given that: a) it is 
conceivable that slit losses could lead to an underestimate of the flux
in a spectroscopic observation, but it is difficult to imagine what 
would lead to an overestimate. Indeed, from  analysis of the FOS target
acquisition records, Maoz et al. (1998) noted that NGC~404 had been 
at the edges of its peak-up scans, possibly leading to some
light loss; b) The HST/FOS 
measurements by Maoz et al. (1998)
for other objects (e.g., NGC~4569), taken with the same setup,
 do agree well with the new measurements, arguing against systematic 
calibration problems. Furthermore, the 1994 FOS spectroscopy indicated
a UV flux of only 65\% of that obtained with the HST/FOC imaging 
measurement in 1993 
by Maoz et al. (1995; $180\times 10^{-17}~\ergcmsA$ at 2300~\AA). 
These measurements imply that the nucleus has faded by a factor $\sim
3$ at 2500~\AA~ between 1993 and 2002. 
Again, for some objects there is excellent
agreement between the FOC, FOS, and ACS measurements, lending credence 
to this conclusion. 
At most, 60\% of the UV light in the spectrum 
obtained in 1994 was contributed
by an AGN, with the rest coming from young  stars (Maoz et al. 1998), 
and therefore  
it appears
that the AGN component at 2500~\AA\ has faded by a large factor
between 1994 and 2002.

\noindent {\bf NGC~1052} --
This galaxy hosts the archetypical LINER~1 nucleus, which has many AGN 
characteristics: weak broad H$\alpha$ wings, which become dominant 
in polarized light, revealing a ``hidden broad-line region'' 
(Barth et al. 1999a); an ``ionization cone'' (Pogge et al. 2000),
reminiscent of those seen in some Seyfert galaxies, aligned with the 
direction of radio
lobes and X-ray knots (Kadler et al. 2004); 
a variable radio core (Vermeulen et al. 2003); and H$_2$O megamaser 
emission (Claussen et al. 1998). Unfortunately,
in our HST snapshot program this object was imaged in the UV only once.
Nevertheless, in 2002 the  UV flux was at half its level in the 
2200~\AA~ region of a 1997 HST/FOS spectrum (Gabel et al. 2000), as
measured by Pogge et al. (2000; $15\times 10^{-17}~\ergcmsA$).
As argued above for NGC~404, the sense of the difference (a lower flux
in the imaging observation), and the reliability of the FOS calibration for
other objects argue that this factor of 2 decline is real. If so, there is a 
significant AGN contribution to the UV light of this nucleus.  
 
\noindent {\bf M81} --
This nucleus is formally a Seyfert 1 (Ho et al. 1997a), since its
[OIII]/H$\beta$ ratio is 30\% above the (rather arbitrary) border
between LINERs and Seyferts. This is close enough that we can safely 
consider it a borderline LINER/Seyfert case.
It has numerous AGN features, including a variable 
and double-peaked broad 
Balmer line component (Bower et al. 1996; Ho et al. 1996), 
a broad-line AGN-like UV spectrum
(Ho et al. 1996; Maoz et al. 1998), and at VLBI resolution, a stationary 
radio core and a one-sided variable jet 
(Bietenholz et al. 2000).
In our current observations, M81 was imaged at five epochs. At 
four of the epochs, the nucleus displays little or no variations in either 
F250W or F330W, with amplitudes of variation 
limited to $<2-3\%$. The flux level 
at 2500\AA~, $200\times 10^{-17}~\ergcmsA$, is similar to
the one measured  by Maoz et al. (1998) at 1500~\AA~ 
in the 1993 HST/FOS spectrum of Ho et al. (1996)
  -- $150\times 10^{-17}~\ergcmsA$. [From an analysis of the FOS target
acquisition records, Maoz et al. (1998) deduced that M81 was 
located at the edges of its peak-up scans, possibly leading to some
light loss.]  The ACS-measured flux at 2500~\AA~ in these four epochs
is also the same as
the 2200~\AA~ flux estimated by Maoz et al. (1998) by extrapolating
the 1996 WFPC2 measurement at $\sim 1600$~\AA~ by Devereux et al. (1997). 
However, the nucleus brightened by $\sim 9\%$ in both F250W and F330W filters
at one epoch, February 2, 2003. The correlated variation in the fluxes
in the two bands, and the lack of anything suspect on that date (e.g., the
nucleus appears unresolved, just as at other epochs), favor that 
this variation is real. Therefore, of order 10\% or more of the 2500~\AA~
and 3300~\AA~ UV continua in this object are nonstellar in nature. M81 is
the bluest object in our sample in terms of 
$f_{\lambda}$(F250W)/$f_{\lambda}$(F330W) color.     
 
\noindent {\bf NGC~3368} -- There is no significant variation in either UV band
between the two epochs at which this LINER~2 was observed.
However, the 2500~\AA~ flux is a factor of 4.5 higher than the 2200~\AA~
flux measured in 1993 with HST/FOC by Maoz et al. (1996;
$5\times 10^{-17}~\ergcmsA$). The large
amplitude of this long-term variation makes it credible, despite the 
difficulties of photometry with the FOC. Note that for the 
nucleus of NGC~3486, which 
had a 2200~\AA~ flux in 1993
($10\times 10^{-17}~\ergcmsA$; Maoz et al. 1996) comparable to that
of NGC~3368, there is good agreement, to 8\%, between the old FOC
measurement  
and the current ACS measurement. 
Thus, NGC~3368 appears to be another
LINER~2 in which the UV is dominated by AGN emission, despite the fact that 
no other AGN features (e.g., a radio core; Nagar et al. 2002) 
have been detected to date.
 
\noindent {\bf NGC~3486} -- This nucleus is one of three objects in our sample
(the others are NGC~4569 and NGC~5055) that show no significant 
short-term or long-term variations. It is a relatively high-ionization
object, borderline between LINERs and Seyferts, and
which Ho et al. (1997) have actually classified as a Seyfert 2 (its
[OIII]/H$\beta$ ratio is 40\% above the formal LINER/Seyfert border).
Its non-variability may be related to its class, as the UV continuum 
of Seyfert 2's is dominated by either scattered nuclear light or
starlight (Gonz\'alez-Delgado et al. 1998).  
 The 2500~\AA~ flux level,
$10.8\times 10^{-17}~\ergcmsA$, is in excellent
agreement with the 2200~\AA~
level in 1993, $10\times 10^{-16}~\ergcmsA$,
measured with the HST/FOC by Maoz et al. (1996).
Since this is another object without other AGN features (e.g., at a
resolution of $1''$, no radio 
core at 6~cm and 20~cm was detected to a 
$3\sigma$ limit of 0.12~mJy~beam$^{-1}$ by Ho \& Ulvestad 2001), 
it is tempting to label it as a non-AGN LINER. Note, however,
that we imaged it on only two, closely spaced (by 1 month), epochs.
For comparison, M81 and M87, which are clearly AGNs with variable UV flux, 
were also near their ``historical'' UV level in the present observations
and were constant in our two closely spaced epochs (for M87)
or in four out of five epochs (for M81). Thus, detection of
short-term variability in NGC~3486 might
 have been possible with better temporal sampling. 
 
\noindent {\bf NGC~3642} -- The 8\% peak-to-peak amplitude variations
in F250W of this
radio-undetected (Nagar et al. 2000) 
LINER~1 are significant at $>98\%$ confidence, based on $\chi^2$.
Fluctuations seen in the F330W band are not significant, for our assumed
photometric uncertainty, although the sense of these fluctuations is correlated
with those in F250W, suggesting they may also be real. The 30\% increase
in 2500~\AA~ flux compared to the 2200~\AA~ WFPC2 measurement in 1994 
by Barth et al. 
(1998; $19\times 10^{-17}~\ergcmsA$) 
is not obviously significant, given the different bandpasses and
the UV sensitivity fluctuations of WFPC2. The small, but significant, F250W
variations show that at least a fraction of the UV flux is nonstellar. 
Interestingly, Komossa et al. (1999) found no evidence in this galaxy for X-ray
variations on short (5 months) or long (years) timescales.  

\noindent {\bf NGC~3998} -- This variable radio-cored (Filho et
 al. 2002)
 LINER~1 displayed  a monotonic 20\%
decline in UV flux in both bands, F250W and F330W, over the 11 months
we observed it. On long timescales, 
its mean 2500~\AA\ flux level in 2003 
was about 5 times lower than reported by Fabbiano et al. (1994; 
$10^{-14}~\ergcmsA$) at 1740~\AA~ in 1992, based on FOC measurements. 
The large amplitude of this long-term variation makes it credible, despite
the different bandpasses and the problems with FOC linearity and dynamic
range. Thus, nonstellar light has dominated the UV output of the nucleus 
over the past decade,
and likely still contributes a significant or dominant fraction.
 
\noindent {\bf NGC~4203} -- This nucleus is a LINER~1 having a
double-peaked H$\alpha$ profile (Shields et al. 2000)
and a variable radio core (Nagar et al. 2002). The radio core  
remains unresolved at the milli-arcsecond scale, and 
Anderson et al. (2004) have shown that its spectrum is most
consistent with that of a jet pointed within $<45^{\circ}$ to our line
of sight.
At Chandra resolution
its compact X-ray nucleus is embedded in soft diffuse emission of 
$50''$ diameter (Terashima \& Wilson 2003).
Its short-term UV variability was the largest in our sample, with a factor of
1.5 between maximum and minimum in F250W, and 1.4 in F330W, and with
the variations
in the two bands clearly correlated. Almost all of the variation occurred
between the first two epochs, separated by 8 months. The 2500~\AA~ flux
level in 2003 was 3-4 times higher than in the HST/WFPC2 2200~\AA~ measurement
by Barth et al. (1998; $21\times 10^{-17}~\ergcmsA$), 
obtained in 1994. This long-term variation is likely real,
given its large amplitude, 
despite the different bandpasses
and the UV photometric instability of WFPC2. 
As already noted, this is the only object in which we detect significant 
color changes, presumably because the variation amplitude is large enough
to reveal the color changes. The sense of the color change, as in luminous
AGN (e.g., Giveon et al. 1999), is that
 the nucleus is bluer when it is brighter. 
 
\noindent {\bf NGC~4258} -- 
     The galaxy with the famous masing water disk 
(Watson \& Wallin 1994; Miyoshi et al. 1995), whose Keplerian 
rotation curve gives the second most accurately measured 
central black hole mass (after the Milky Way), has been variably classified
as a LINER~1 or a Seyfert~1.9. In the spectrum of 
Ho et al. 1997a, its emission-line ratio of 
[OIII]/H$\beta=10$, which is 3 times greater than the the formal  border
between LINERs and Seyferts and hence well in the Seyfert domain. 
 Wilkes et al. (1995) and Barth et al. (1999b) showed that the
spectrum in polarized light has emission lines 
that are broader than the lines in the total flux
spectrum. However, 
this is seen not only in the Balmer lines but in most of the forbidden lines
as well, with the width of the lines in the polarized spectrum 
depending on the critical density
of the transition. The phenomenon is thus different 
from that of the hidden broad-line regions
revealed in polarized light in some Seyfert~2 galaxies. 
Possible explanations for the effect are the presence of 
a structure that obscures and polarizes the inner parts of the
     narrow-line region (Barth et al. 1999b),
or a broadening of the lines due to the impact of the jet
on the emission line gas (see, e.g., Wilson et al. 2000).
 
Our measurements indicate significant fluctuations in nuclear flux, 
with a peak-to-peak 
amplitude of 16\% in F250W and 8\% in F330W. Contrary to the other objects
with significant variations detected in both bands, the variations in this 
galaxy are not perfectly correlated between the bands, particularly 
on the third epoch. 
Pogge et al. (2000) estimated the 2200~\AA~ flux from a F218W 
WFPC2 image from 1997,
at $7\times 10^{-17}~\ergcmsA$. Given the photometric UV instability of
WFPC2 (for which Pogge et al. made no correction), and the significant
red leak in the WFPC2+F220W configuration, plus the fact that this is 
the reddest nucleus in our sample in  terms of 
$f_{\lambda}$(F250W)/$f_{\lambda}$(F330W) color, the WFPC2 flux level is
consistent with the mean ACS F250W level, $5.2\times
10^{-17}~\ergcmsA$. Thus, there is no evidence for
long-term variation between 1997 and 2003. As already noted,
this object is also anomalous in the sense that it has some
type-1 characteristics but its UV color is red. Pogge et 
al. (2000) argued that it likely undergoes foreground reddening, perhaps by the
dust in the molecular gas disk that produces this galaxy's observed
 water masers.
  
\noindent {\bf M87} -- Although a LINER~2 (Ho et al. 1997a), with no detected
broad component to its Balmer lines, M87 has numerous AGN features,
most notably its prominent radio-through-X-ray jet. We obtained only two epochs
on this object, separated by 40 days, and showing only marginally significant
(94\% confidence, based on $\chi^2$) variation in F250W and no significant
variation in F330W. However, as noted above, Perlman et al. (2003) monitored
M87 with ACS/HRC and the F220W filter at five epochs between November 2002
and May 2003. Their light curve, which includes the two epochs obtained
by us, shows clear nuclear variability, with about 20\% peak-to-peak 
amplitude. The 2500~\AA~ flux is about half that measured by 
Maoz et al. (1996; $1\times 10^{-15}~\ergcmsA$) using an archival HST/FOC 
image from 1991.
There is thus no doubt that AGN emission contributes significantly
to the nuclear UV flux from this object. 
The radio-to-X-ray emission in the jet is
certainly synchrotron. Hence it is likely that the nuclear UV
emission is also synchrotron emission from the jet, 
or at least has a very strong jet
contribution.
 
\noindent {\bf NGC~4552} -- As already noted above, 
the classification of this nucleus is uncertain.
Ho et al. (1997a) labeled it a type-2 object, on the border between
LINERs and H~II regions, with no detected broad H$\alpha$ component.
 Cappellari et al. (1999), analyzing an HST/FOS spectrum,
found the narrow-line ratios were borderline between LINERs and Seyferts,
and detected a broad component in both the Balmer lines and in the
forbidden lines. It is therefore unclear whether this
galaxy is more akin to type 1s or type 2s.
Its blue $f_{\lambda}$(F250W)/$f_{\lambda}$(F330W) color is 
certainly reminiscent of type 1s.
Between the two epochs at which we observed it, the nucleus brightened by 20\% 
in both UV bands. This confirms the previous reports of long-term 
variability in this object by  Cappellari et al. (1999; see \S1). 
The mean of the two 2500~\AA~ data points,
$2\times 10^{-17}~\ergcmsA$, is the same
as was measured with HST/FOS in 1996 by Cappellari et al. (1999), and close 
to their HST/FOC measurements obtained in 1993, 
$1.5\times 10^{-17}~\ergcmsA$ (in F220W) and $1.8\times 10^{-17}~\ergcmsA$
(in F275W). 
 
\noindent {\bf NGC~4569} -- A ``transition'' nucleus between an H~II nucleus
and a LINER~2 (Ho et al. 1997a), this object is one of three that showed
no variations in either UV band. Its constancy is not surprising,
since Maoz et al. (1998) showed, based on its UV spectrum,
 that at least 80\% of the UV flux is stellar. (Of course, this left 
room for a 20\% AGN component, but we have found no evidence for such a 
component in the present experiment). The mean 2500~\AA~ flux level we measure
with ACS, $(0.992\pm 0.005)\times 10^{-14}~\ergcmsA$, is in excellent agreement
with previous measurements at $\sim 2200$~\AA~ -- 
FOS: $1.05\times 10^{-14}~\ergcmsA$
(Maoz et al. 1998) -- FOC: $1.0\times 10^{-14}~\ergcmsA$
(Maoz et al. 1995) -- and WFPC2:  $1.1\times 10^{-14}~\ergcmsA$
(Barth et al. 1998). This agreement lends credence to the detection
of long-term variations in similar comparisons for the other objects
in our sample.
 
\noindent {\bf NGC~4579} -- This nucleus, 
classified by Ho et al. 
(1997a) as a transition case between a LINER~1 and a Seyfert 1.9,
has a radio core that 
remains unresolved at the milli-arcsecond scale, 
with a  spectrum that is most
consistent with that of a jet pointed within $<40^{\circ}$ to our line
of sight (Anderson et al. 2004).
Our new images show
in sharp detail the disk/spiral arm around the point-like central source,
already visible in the UV image by Maoz et al. (1995; 1996). Its
X-ray morphology, as viewed with Chandra, is a very bright nucleus
surrounded by soft diffuse emission from the circumnuclear
star-forming ring (Eracleous et al. 2002; 
Terashima \& Wilson 2003). Barth et al. 
(1996)  noted a factor 3 fading in UV flux between the 1993
FOC images of Maoz et al. (1995) and their own FOS observations in 1994,
obtained 19 months later.
This apparent large variation was further studied by Maoz et al. (1998),
who again concluded it is likely real. Our current data leaves no doubt 
regarding the large variability of this object. Between the two epochs 
at which it was observed, separated by less than a month, it brightened by 7\% 
in both UV bands. The mean flux level at 2500~\AA~ in 2003,
$61\times 10^{-17}~\ergcmsA$, is between, but significantly
different from, the two
previous measurements at $\sim 2200$~\AA: 
$33\times 10^{-17}~\ergcmsA$
(Maoz et al. 1998) and $110\times 10^{-17}~\ergcmsA$
(Maoz et al. 1995).     
 
\noindent {\bf NGC~4594} -- The LINER~2 
nucleus of the Sombrero galaxy appears unresolved and isolated
at 2500~\AA, as previously seen at 2400~\AA~ by Crane et al. (1994).
The nucleus shows large short-term
variations, with 20\% peak-to-peak amplitude in F250W 
and 11\% in F330W, which are
clearly correlated in the two bands. The 2500~\AA~ flux is at 2/3 of its level
measured with the FOS in 1995 (Nicholson et al. 1998; 
Maoz et al. 1998), a change that is 
likely real.  
 
\noindent {\bf NGC~4736} -- This ringed Sab galaxy has a LINER~2 nucleus,
an exceptionally 
bright central surface brightness in the optical and infrared, 
and is thought to be an aging starburst (e.g., Waller et al. 2001,
 and references therein). VLA measurements with resolution $0\farcs 15$
reveal an unresolved nuclear source with flux 1.7~mJy at 2~cm
(Nagar et al. 2004). 
There is a $6''$ offset between
the position of the radio core, and the position of the 
 optical nucleus reported by  Cotton et al. (1999), 
based on a measurement from the digitized Palomar Sky Survey (POSS). 
However, the optical position of the nucleus given by  Cotton et al. (1999)
is erroneous. We find agreement to better than 1 arcsecond between
the 15~GHz radio core position, the radio core position reported
by the VLA-FIRST survey (Becker et al. 1995), 
the optical position we measure from digitized
POSS plates, and the near-IR nuclear position from the 2MASS survey
(Jarrett et al. 2003). 
From all of these, the J2000 coordinates of the radio-through-IR nucleus are
RA:~$12^h 50^m 53^{s}_{.}06$~DEC:~$41^{\circ}07'12\farcs 7$. 

Maoz et al. (1995) noted a second UV source, of
 brightness comparable to the nuclear one, $2\farcs 5$ to the north of the
 nucleus (at position angle $-2.4^{\circ}$). 
The distance to this galaxy is $\sim 4.9$~Mpc, taking the
 mean between 5.2~Mpc found by Tonry et al. (2001) from surface
 brightness fluctuations, and 4.66~Mpc found by Karachentsev et al. 
(2003) based
 on the tip of the red-giant branch. At this distance, the separation
 between the two UV sources is 60~pc.
Maoz et al. (1995) speculated that the off-nuclear source, 
which we designate NGC~4736b, 
could be the active nucleus of a galaxy  that
had merged with NGC~4736. Perhaps this merger triggered the past starburst 
and the peculiar morphological and kinematic features observed in this
galaxy. 

Maoz et al. (1996) measured in 1993 the FOC 2200~\AA~
fluxes of the nuclear UV source, which we designate NGC~4736, and of
NGC~4736b, as $19\times 10^{-17}~\ergcmsA$ and $26\times 10^{-17}~\ergcmsA$,
respectively. NGC~4736    
varied slightly during the current program, 
and only in the last measurement in F250W.
Nevertheless, NGC~4736 was 2.5 times brighter in 2003 than in 1993. 
 Given the 
large amplitude of the change in NGC~4736, 
and the reliability of the FOC photometry
for several apparently non-variable objects, we believe the factor 2.5
brightening of the nucleus is real.  
As for NGC~4736b, its mean flux level is similar to that measured in 1993
by Maoz et al. (1996). However, in the present measurements 
this off-nuclear source shows significant,
correlated F250W and F330W fluctuations with peak-to-peak amplitudes of 5\%.
We note that, due to the large brightnening of the nuclear UV source, 
NGC~4736b was 30\% brighter
than NGC~4736 in 1993, while in 2003 
the nucleus was 70\% brighter than NGC~4736b.
We conclude that we have detected long-term UV variations
in the nucleus of NGC~4736, and short-term variations
in the off-nuclear source NGC~4736b, indicating
significant nonstellar contributions to the UV fluxes of both sources.
We also note that the two sources have quite different UV colors,
with the nuclear source being red and NGC~4736b being blue.
 
We have searched the VLA archive for deep radio observations of 
this galaxy, to see if object b has, at any time, shown up as a radio
source. We have found seven different VLA epochs  of observation 
of NGC~4736 between December 1984 and
February 2000, with useful data at 2 to 20~cm. In addition,
NGC~4736 was monitored at eight epochs at 3.5~cm between June and October 2003
by K\"ording et al. (2005). In none of these radio
maps do we see a signal at the position of the northern, off-nuclear,
source. The deeper archival images are from December 1984 and January  1985, 
from which we can 
put a $3\sigma$  upper limit of  $150~\mu$Jy on the radio flux at
20~cm and 6~cm, respectively, and from December 1988, which 
puts a $3\sigma$  upper limit of  $300~\mu$Jy  at 20~cm. 
The  K\"ording et al. (2005) data from 2003
place a $3\sigma$  upper limit of  $150~\mu$Jy  at each epoch,
or an upper limit $60~\mu$Jy  from the eight epochs combined.
Interestingly, in the higher-resolution observations among these
datasets, the nuclear source is resolved into two
sources of comparable flux, separated by $0\farcs 99$,
at position angle $-49^{\circ}$. This second source, which is reported
and described in detail by  K\"ording et al. (2005),
may be the result of jet activity in the
nucleus, or some other radio source that is very close to the nucleus.

High spatial resolution {\it Chandra} X-ray images of NGC~4736 were obtained
by Eracleous et al. (2002). They detected an unresolved nuclear source
with a 0.5-8 keV luminosity of  $5.9\times10^{38}$~erg~s$^{-1}$, and
39 other sources, presumably X-ray binaries and supernova remnants,
distributed around the nucleus. However, they did not detect an X-ray
source at the position of NGC~4736b, to a luminosity limit
of $1\times10^{36}$~erg~s$^{-1}$. Interestingly, variability data
shown by Eracleous et al. (2002) suggest that nucleus may be variable
in the X-rays on hour timescales, with a measured excess variance
of $0.06\pm 0.04$. 
  
Alternatively to the binary AGN scenario, the variable off-nuclear UV source
NGC~4736b could be related to jet activity emerging from the nucleus.
For example, the off-nuclear source could be at a location where gas is
heated by beamed radiation, or it could be synchrotron radiation from freshly 
accelerated particles at the end of a jet, with a spectrum that is
hard enough to avoid detecting this knot in radio. Such a scenario
would be reminiscent of NGC~1052, where Kadler et al. (2004) find
spatial offsets between optical, X-ray, and radio knots associated
with the jet.    

To further investigate the nature of the two UV sources
requires high spatial resolution 
optical-UV observations that will elucidate the spectral 
properties of each of the sources. At present, it is not
clear whether the LINER~2 spectrum attributed to the central parts of
this galaxy is emitted
by the nuclear source NGC~4736a, the off-nuclear source NGC~4736b, or both.

\noindent {\bf NGC~5055} -- This transition H~II/LINER~2 nucleus 
is not variable
in our data.
The nuclear UV source is clearly extended, with a FWHM of $0.14''$.
The 2500~\AA~ flux level, $76.4\times 10^{-17}~\ergcmsA$, is
virtually identical to that measured with the FOS in 1996 by 
Maoz et al. (1998), though 25\% less than the 1993 FOC measurement
by Maoz et al. (1995). The latter difference is also consistent
with non-variability, considering
the uncertainties in FOC photometry, which are further complicated
by the extended nature of this source. 
Like the other non-variable LINER~2 in our sample, NGC~4569, the FOS UV
spectrum of this object indicates a $>50\%$ hot-star contribution
to the UV flux (Maoz et al. 1998). Our results suggest that
NGC~5055 is a member of the minority of LINERs whose UV flux
is all or mostly from stars. 
 
\noindent {\bf NGC~6500} -- This LINER~2 has several AGN traits,
including a radio-core (Nagar et al. 2000), and a jet-like linear
structure seen with the VLBA (Falcke et al. 2000)
As seen in Fig.~1, and known from 
previous imaging with WFPC2 by Barth et al.
(1998), NGC~6500 does not have a clear
nuclear UV source. The UV morphology of the central region consists of 
a diffuse central light distribution, on which are superposed
a number of faint sources, some compact and some extended,
within a diameter of $\sim 0\farcs5$. 
It may actually be a ``UV-dark'' LINER (like 75\% of
all LINERs; Maoz et al. 1995; Barth et al. 1998), that happens to 
possess some scattered circumnuclear star formation. Maoz
et al. (1998) noted that its observed UV luminosity at 1300~\AA,
extrapolated to the far UV,  
is insufficient to
power its  H$\alpha$ luminosity. Terashima \& Wilson (2003) found that
the nuclear X-ray emission observed with Chandra is anomalously faint
given the H$\alpha$ luminosity and the X-ray vs. H$\alpha$ correlation
observed in other AGNs, and proposed that the nucleus is heavily 
obscured in X-rays. Our large-aperture
measurements for this galaxy (see \S 2) 
indicate no variability in F330W, but fluctuations 
in F250W that are formally significant at 99\% confidence, and driven
mainly by a 5\% drop at the last epoch. There is nothing suspect with
the data at this last epoch. There is a compact source $9''$
east of the nucleus that appears in the first two epochs and the last
epoch (in the third epoch it is obstructed by the HRC ``occulting
finger''), and which can serve as a local calibrator. 
Its flux is constant to 0.7\%, and its FWHM is similar
at all three epochs.     
However, we are not certain of the reality of the 5\% decline at the
last epoch for a number of reasons:
the large aperture and the diffuse nature of the object increase
 the susceptibility  of the measurement to small fluctuations due to
centering differences. We do not have
an independent photometric stability check for such a diffuse source,
as we do for the compact nuclear sources, 
through monitoring of the star cluster NGC~6681 by
 Boffi et al. (2004) -- see \S3.2. Indeed, the amplitude of the
 decline on the last epoch depends on the choice of region used to 
determine the background level, and for some choices the decline
is only 2\%. The fairly large variation at one
epoch, observed solely in one filter
is contrary to what we have seen in all the other
objects, where large variations are mirrored in the two bands.
We have blinked and compared the images of the four epochs to try to identify
a particular knot in the nuclear region 
that declined in brightness on the last epoch, but have not been able 
to reach a definitive conclusion.  
The 2500~\AA~ flux, $31\times 10^{-17}~\ergcmsA$,
 is similar to that measured by 
Maoz et al. (1998) at 2200~\AA~ from the Barth et al. (1997) 
FOS spectrum taken in  1994,  and to that measured with WFPC2 in 1995 
by Barth et al. (1998). In both cases, the UV flux was 
$27\times 10^{-17}~\ergcmsA$. The difference between our, and these
earlier, measurements is not significant,
especially considering the extended nature of the source.

\section{Discussion and Conclusions}

Our UV monitoring program has revealed, by means of variability, that
 an AGN component contributes to the UV emission of most UV-bright
LINERs. Variability is detected irrespective of spectral type
(1 or 2) and whether or not a nuclear radio source has been detected. 
LINERs are 
present in the majority of massive galaxies, and the true fraction 
of LINER galaxies that have a nuclear UV 
source, after accounting for
extinction, is likely close to unity (Barth et al. 1998; Pogge et al. 2000).
This conclusion 
implies that, not only do most galaxies have central black holes,
but that the black holes are also accreting and emitting in the UV.
The extreme-UV extension of the observed UV, beyond the Lyman limit,
 is the main
ionizing agent in these objects, and it determines the optical line ratios
that define LINERs and distinguish them from other nuclei (e.g., H~II,
Seyfert). Our results provide some of the strongest evidence to date that,
in the majority of cases, a LINER spectrum in fact signals the presence 
of nonstellar activity (i.e., an AGN).  
Our data confirm previous 
reports of large-amplitude UV variability in several LINERs (see \S 1), 
but now with a stable photometric setup, applied systematically to a 
moderate-sized sample.
 
We have identified only three galaxies with UV nuclei 
that show neither short-term ($\lesssim 1$~yr) nor long-term 
($\gtrsim 1$~yr) variations. In all three cases,
there is previous evidence that stars dominate the UV emission --
based on UV spectra in NGC~4569 and NGC~5055 (Maoz et al. 1998), and 
on the Seyfert 2 classification (Ho et al. 1997a) in the case of NGC~3486,
combined with the fact that the observed UV emission in
Seyfert 2s often comes from stars (Gonz\'alez-Delgado et al. 1998).  
However, even in these three
cases, it
is possible that UV variability would be detected in an experiment with
denser or longer-term sampling. 

 Interestingly, 
all three galaxies without detected UV variations
have no detected radio cores, to 1~mJy sensitivity for NGC~4569
and NGC~5055 (Nagar et
al. 2000) and to 0.12~mJy sensitivity for NGC~3486 (Ho \& Ulvestad
2001). 
Conversely, all the LINERs that do have detected radio
cores have variable UV nuclei. Of course, there are three galaxies
(NGC~404, NGC~3368, and NGC~3642) with no detected radio core that
do display UV variations, but radio cores may be revealed by more sensitive 
observations of these
objects. If deeper radio observations revealed cores in the three 
UV-variable LINERs, but not in the three UV-stable LINERs, a perfect
correspondence would exist between the presence of radio cores and   
UV variability.

Could stars produce the observed UV variations? The
2500~\AA~ luminosities of the variable nuclei in our sample, 
as seen in Table~2, are distributed more or less
evenly in the range  $L_{\lambda}(2500~{\rm \AA})\sim
10^{35.6}-10^{37.7}~\ergs~{\rm \AA}^{-1}$. 
The most luminous ``normal'' stars are blue 
supergiants, among which those with effective temperatures of
$1-3\times 10^4~K$ (spectral classes B0--A0) have the highest near-UV 
luminosities. 
 Bresolin et al. (2004)
recently monitored 70 blue supergiants in the galaxy NGC~300 
over a 5-month period, and found 15
of them to be variable, with $V$-band amplitudes of $8-23\%$. 
The mean $V$ band absolute magnitude of these variable supergiants is 
$M_V=-7.3$, and the most luminous one (of spectral type B9) has
$M_V=-8.7$.
Using spectral models by Kurucz for B9 supergiants 
to obtain the flux ratio between the $V$ band and 2500~\AA, the
UV luminosities for the typical and for the most luminous
supergiants are   
$L_{\lambda}(2500~{\rm \AA})=
10^{34.6}~\ergs~{\rm \AA}^{-1}$ and $L_{\lambda}(2500~{\rm \AA})=
10^{35.2}~\ergs~{\rm \AA}^{-1}$, respectively. Thus,
even the least luminous galactic nuclei in our sample have luminosities about
an order of magnitude larger than typical blue supergiants, but
short-term variability amplitudes of 15--20\%, comparable to those of
the most variable supergiants. Furthermore, the large amplitude
variations
we see in many of the LINERs are not expected in supergiants.
Therefore, individual supergiants in
the galactic nuclei are
not plausible candidates for producing the observed level of UV variability   
in the LINERs.

Stars more luminous than blue supergiants, by an order of magnitude,
 do exist --
Wolf-Rayet (WR) stars (e.g., Conti 2000) and
luminous blue variables
(LBVs; e.g., Humphreys \& Davidson 1994) with bolometric luminosities 
up to $L_{bol}\sim 10^{40}~\ergs$. These are massive stars, nearing
 the end of their evolution and 
 radiating near the Eddington limit. LBV variation 
amplitudes on timescales of months
are $\lesssim 10\%$ (Humphreys \& Davidson 1994), whereas
WN8 stars, which are the most variable among WRs, sometimes
vary in the optical by a few percent on week-to-month timescales
(Marchenko et al. 1998). Based on luminosity 
and variation amplitude alone, individual LBVs and WR stars
could perhaps produce the UV variations in part of the lower-luminosity half of
our sample. However,  
 the variations seen in the 
most luminous objects in our sample certainly cannot
be explained in this way, and, based on continuity, one can then argue
that stars are not the source of variations in any of the LINERs.
Nevertheless, we cannot rule out the possibility that individual evolved
stars dominate the light output in a few of the low-luminosity
objects. For example, spectroscopy of NGC~4736b, with a luminosity 
of $L_{\lambda}(2500~{\rm \AA})=
10^{36}~\ergs~{\rm \AA}^{-1}$, could reveal if it is a LBV or WR
star, rather than a second merging nucleus. 
 
Returning to the AGN interpretation, 
the {\it variable} flux that we have measured in each UV band,
in the absence of any extinction corrections, provides
a firm {\it lower} limit to the AGN flux in that band. This observed 
lower limit can be used to test accretion models for each of these
low-luminosity AGNs. 
It can also be argued that the {\it total} UV flux provides
 an {\it upper} limit on the UV emission, but the sensitivity of the UV to
uncertain extinction corrections  makes such an upper limit less
robust. The VLBA images of Falcke et al. (2000) and 
Nagar et al. (2002) have shown that at least some of the radio emission in 
LINERs is contributed by jets, rather than by an actual accretion flow. 
Anderson et al. (2004) have obtained multifrequency VLBA
spectra for the unresolved milli-arcsecond core in six low-luminosity
AGNs (including three LINERs, two of which, NGC~4203 and NGC~4579,
are in our sample). They showed that the radio spectra are
inconsistent with expectations from accretion flows (cf. Nagar et al. 2001), 
but that the spectra, luminosity, and size limits are consistent with 
 emission from jets that are pointed toward us to within 
$\lesssim 50^{\circ}$ . The
observed radio flux must therefore constitute only an upper limit
on radio emission from the accretion flow itself. 

Strictly speaking,
X-ray data,
too, provide only upper limits to the flux from the accretion flow, since
even with the excellent spatial resolution of {\it Chandra}, 
non-nuclear X-ray sources (low-mass X-ray binaries, 
supernova remnants, diffuse emission) can be 
included in the beam, and may contribute to the X-ray flux. 
As an extreme example, in M32 the nuclear X-ray source
produces only 1\% of the total X-ray luminosity within a radius of
30~pc of the nucleus (Ho et al. 2003), yet this is the area covered by a 
$\sim 1''$-diameter beam at 10~Mpc, the typical distance to a galaxy in
our sample. Nevertheless, the excellent astrometric agreement 
($\lesssim 0\farcs
5$) beween X-ray and radio positions in low-luminosity AGNs 
(e.g., Terashima \& Wilson 2003), and the absence of close 
by ($<$ few arcseconds) X-ray sources indicates that, in most such AGNs, 
the X-rays originate from the active nucleus.

 At optical and IR wavelengths, the nuclear
emission cannot be detected, at present, in the face of the bright stellar
 backgrounds, and in the extreme-UV only indirect, model-dependent estimates
of the SED can be obtained by attempting to reproduce the UV-through-IR 
emission line fluxes and ratios. The current flux limits (lower limits
to the emission from the accretion flow in the UV, from our
present results, and upper limits in the radio and X-rays, 
from previous observations)
thus provide a potentially powerful test of accretion models. 
The observed SED can similarly be compared to the predictions of jet
models. In this case, the UV emission should be connected to emission
at other wavelengths and to the assumed black-hole mass by a relation 
analogous to the radio/X-ray correlation seen in low-accretion-rate
black holes (Merloni et al. 2003; Falcke et al. 2004) 
We intend to carry out such comparisons to models in a future paper.

Our data for NGC~4736 have revealed the first example of a variable
off-nuclear UV source, giving new grounds to previous speculation 
(Maoz et al. 1995, 1996) that
this is a ``wandering black hole'' from the nucleus of another galaxy
that has recently merged with this one. This hypothesis 
(and the alternative, that it is an individual WR or LBV star, see
above) can be tested
with straightforward high-spatial-resolution observations in radio,
optical, UV, and X-ray bands. 
The observational evidence for the existence of systems of
massive black hole pairs has been reviewed recently by Komossa (2003). 
The best current candidate for a double AGN 
is NGC~6240, a relatively
nearby (redshift $z=0.024$) ultraluminous infrared galaxy. 
Both nuclei of NGC~6240 are
probably LINERs (Raffanelli et al. 1997), both are compact radio
sources
at 1.4 and 5~GHz (Colbert et al. 1994; 
Gallimore \& Beswick 2004), and both emit hard X-ray
continua and Fe~K$\alpha$ lines (Komossa et al. 2003). The $1.5''$
separation of the nuclei in NGC~6240, for an assumed distance of
100~Mpc, corresponds to a projected
physical scale of $\sim 700$~pc, compared to only 60~pc between the 
possible double nuclei of NGC~4736. Another, somewhat more ambiguous,   
case is NGC~3256, a merging galaxy system at a distance of about 40~Mpc. 
In this case,
two nuclear sources, with a projected separation of $5\farcs 2$, i.e.,
about 1~kpc, are detected in near-infrared and X-rays (Lira et al. 2002) and
in radio (Neff et al. 2003), with a radio-to-X-ray flux ratio that is
characteristic of low-luminosity AGN.    
In the central CD galaxy of the cluster Abell 400, the twin-jetted 
double radio source 3C~75 (Owen et al. 1985), 
is a spectacular example of a
binary AGN, though with a rather large separation of 7~kpc.  
Identification and study of new examples of binary AGN,   
especially as nearby as NGC~4736,
can shed light on the issue of the rate of coalescence of supermassive black 
holes (e.g., Quinlan \& Hernquist 1997).

We have found a possible UV-color-based indicator of whether a LINER
is a type-1 or type-2 object. If confirmed, this would be the first
LINER property that is found to be linked with the presence or absence
of a broad Balmer emission line component. As we have argued above, 
reddening by dust is not obviously 
the mechanism behind the suggested trend --
the factor $\sim 2$ difference in 
the UV color ratio of type 1s and type 2s, if produced by a dust
screen,
would correspond to a mere factor of 2.4 increase in the 
 extinction of a hypothetical
broad H$\alpha$ line in LINER 2s.
Instead, we argue that the instrinsic
color of the UV continuum is related to the existence or the visibility 
of a BLR. 
For example, the physical conditions under which a BLR can form could 
depend on the present accretion mode, which might be reflected in the UV color.
Nevertheless, as we have pointed out, some combination of dust,
geometry, optical depth and selection effects may be, after all, behind the
observed trend of UV color with LINER spectral type. Furthermore, 
the trend itself  needs to be
confirmed with a larger sample. 
This is not a simple task, since we have imaged all known
LINERs that have a compact nucleus in the space-UV. A larger sample
of such objects could
be assembled by means of UV imaging (e.g., with GALEX) and
subsequent optical spectroscopic classification to identify the LINERs. 
UV imaging need not necessarily be from space -- the UV nuclei of our
current sample are prominent in the F330W band, so such objects could
potentially be indentified by ground-based 
observations near the atmospheric UV cutoff.
Larger samples of LINERs 
could also be produced by 
studying a fainter sample of galaxies 
than that of Ho et al. (1997a), based, for example,
on the Sloan Digital Sky Survey, or by surveying the Southern hemisphere.

After we submitted this paper, Totani et al. (2005) reported
discovering optical nuclear variability in a ``blind'' variability
search among $\sim 1000$ massive galaxies at redshifts $z\sim
0.3-0.4$. They found six nuclei with estimated variability amplitudes
of order unity over a one-month timescale, with marginal evidence for
day-to-day variations. Spectroscopy of one of the six variable nuclei 
revealed a LINER spectrum at $z=0.33$, with an H$\alpha$ flux implying
a specific optical luminosity $L_{\lambda}=2\times 
10^{37}$~erg~s$^{-1}$~\AA$^{-1}$, quite similar to the objects
studied in this paper. It appears plausible that Totani et al. (2005)
have discovered, at $z\sim 0.3$,
 the large-amplitude-variability tail of the variations
we have found in nearby LINERs. Assuming that of order one-half of
early-type galaxies are LINERs (Ho et al. 1997a), and that one-fourth
of LINERs have unobscured optical/UV continua (Maoz et al. 1995; Barth
et al. 1998), there would be in the data of Totani et al. 
of order 100 galaxies of the type we have studied here. Since only one
galaxy among the 15 that we monitored with HST on month-long time
scales showed a variability amplitude of order unity (NGC~4203, which
varied by $\sim 40$\%), it is to be expected that Totani et al. would
detect about six such cases.   

\acknowledgments
We thank Tricia Royle for her expert assistance in the
implementation  of the observing program, and Luis Ho, Eva Schinnerer,
Amiel Sternberg, Joe Shields, and an anonymous referee, 
for useful advice and input.
This work was funded in part by grant GO-9454 from the Space Telescope
Science Institute, which is operated by AURA, Inc., under NASA contract NAS
5-26555. 
This research has made use of the NASA/IPAC Extragalactic Database
(NED) which is operated by the JPL, 
Caltech, under contract with NASA.
This publication also makes use of data products from the 
Two Micron All Sky Survey, which is a joint project of the University
of Massachusetts and IPAC/Caltech, funded by NASA and the NSF.

\begin{deluxetable}{lrrrrrrrr}
\tablecolumns{9}
\tabletypesize{\scriptsize}
\tablewidth{0pt}
\tablecaption{Observations and Photometry
\label{agndata1}}
\tablehead{
\colhead{Object} &
\colhead{Exp.} &
\colhead{Exp.} &
\colhead{UT-Date} &
\colhead{M.J.D.} &
\colhead{$f_\lambda$} &
\colhead{$\sigma$} &
\colhead{$f_\lambda$} &
\colhead{$\sigma$} \\
\colhead{} &
\colhead{F250W} &
\colhead{F330W} &
\colhead{} &
\colhead{} &
\colhead{F250W} &
\colhead{} &
\colhead{F330W} &
\colhead{}\\
\colhead{(1)} &
\colhead{(2)} &
\colhead{(3)} &
\colhead{(4)} &
\colhead{(5)} &
\colhead{(6)} &
\colhead{(7)} &
\colhead{(8)} &
\colhead{(9)} 
 }
\startdata
NGC~404 & 300& 300& 2002-10-28 & 175.1 &  73.57 &    0.86 &  85.27  &   0.90   \\
NGC~1052 & 300& 300& 2002-10-18 & 165.4 &   7.65 &    0.27 &   8.97  &   0.16   \\
 M81     & 300& 300& 2002-08-05 &  91.5 & 198.35 &    2.09 & 126.51  &   1.31   \\
         &   &   & 2002-11-27 & 205.9 & 199.07 &    2.10 & 125.77  &   1.31   \\
         &   &   & 2003-02-02 & 272.9 & 214.31 &    2.25 & 140.01  &   1.45   \\
         &   &   & 2003-04-07 & 336.5 & 200.98 &    2.12 & 130.01  &   1.35   \\
         &   &   & 2003-06-12 & 402.3 & 202.54 &    2.13 & 128.93  &   1.34   \\
NGC~3368 & 600& 300& 2002-10-25 & 172.1 &  22.53 &    0.30 &  30.00  &   0.36   \\
         &   &   & 2003-05-10 & 369.0 &  21.99 &    0.29 &  29.63  &   0.35   \\
NGC~3486 & 600& 300& 2003-04-06 & 335.7 &  10.70 &    0.19 &  18.11  &   0.24   \\
         &   &   & 2003-05-10 & 369.4 &  10.82 &    0.19 &  18.11  &   0.24   \\
         &\nodata&1200& 2003-06-03 & 393.5 & \nodata   &   \nodata   &  18.24  &   0.20   \\ 
NGC~3642 & 300& 300& 2002-10-02 & 149.3 &  25.28 &    0.41 &  22.19  &   0.28   \\
         &   &   & 2003-01-20 & 259.3 &  24.50 &    0.41 &  22.24  &   0.28   \\
         &   &   & 2003-04-12 & 341.3 &  24.62 &    0.42 &  22.34  &   0.28   \\
         &   &   & 2003-05-26 & 385.5 &  23.55 &    0.40 &  21.74  &   0.28   \\
NGC~3998 &  60&  60& 2002-07-01 &  56.0 & 220.46 &    2.91 & 168.38  &   1.96   \\
         &   &   & 2002-11-13 & 191.3 & 210.84 &    2.82 & 162.35  &   1.90   \\
         &   &   & 2003-03-05 & 303.3 & 193.63 &    2.66 & 145.96  &   1.74   \\
         &   &   & 2003-04-05 & 334.4 & 185.68 &    2.58 & 146.38  &   1.75   \\
         &   &   & 2003-05-29 & 388.4 & 182.69 &    2.56 & 140.14  &   1.69   \\
NGC~4203 & 300& 300& 2002-07-03 &  58.7 &  77.51 &    0.90 &  46.45  &   0.52   \\
         &   &   & 2003-03-05 & 303.0 &  54.98 &    0.68 &  34.58  &   0.40   \\
         &   &   & 2003-04-07 & 336.8 &  54.20 &    0.68 &  35.27  &   0.41   \\
         &   &   & 2003-04-20 & 349.0 &  50.68 &    0.64 &  33.69  &   0.39   \\
         &   &   & 2003-06-13 & 403.8 &  53.79 &    0.67 &  35.95  &   0.41   \\
NGC~4258 & 600& 300& 2002-07-06 &  61.1 &   5.02 &    0.15 &  11.58  &   0.19   \\
         &   &   & 2002-10-21 & 168.3 &   4.78 &    0.14 &  10.79  &   0.18   \\
         &\nodata&1140& 2002-12-07 & 215.2 & \nodata   &   \nodata   &  10.82  &   0.12   \\
         &   & 300& 2003-04-17 & 346.5 &   5.56 &    0.15 &  10.82  &   0.18   \\
         &   &   & 2003-06-28 & 418.1 &   5.26 &    0.15 &  11.58  &   0.19   \\ 
 M87     & 300& 300& 2003-03-31 & 329.8 &  56.24 &    0.70 &  48.64  &   0.54   \\
         &   &   & 2003-05-10 & 369.0 &  54.44 &    0.68 &  48.19  &   0.53   \\
NGC~4552 &1500& 750& 2003-03-23 & 321.7 &   1.79 &    0.06 &   1.34  &   0.05   \\
         &   &   & 2003-06-03 & 393.2 &   2.15 &    0.06 &   1.63  &   0.05   \\
NGC~4569 &  60&  60& 2002-07-03 &  58.1 & 999.53 &   10.55 & 841.30  &   8.65   \\
         &   &   & 2003-02-02 & 272.6 & 983.03 &   10.38 & 840.11  &   8.63   \\
         &   &   & 2003-03-31 & 329.8 & 988.35 &   10.44 & 832.59  &   8.56   \\
         &   &   & 2003-04-29 & 358.0 & 999.23 &   10.54 & 834.64  &   8.58   \\
NGC~4579 & 300& 300& 2003-03-17 & 315.2 &  59.05 &    0.72 &  40.34  &   0.46   \\
         &   &   & 2003-04-12 & 341.9 &  63.17 &    0.77 &  43.42  &   0.49   \\
NGC~4594 & 300& 300& 2003-03-24 & 322.5 &   6.93 &    0.27 &  14.37  &   0.21   \\
         &   &   & 2003-05-05 & 364.9 &   8.43 &    0.28 &  16.13  &   0.23   \\
         &   &   & 2003-06-09 & 399.4 &   7.05 &    0.27 &  15.37  &   0.22   \\
NGC~4736 & 300& 300& 2003-03-20 & 318.6 &  47.63 &    0.61 &  72.87  &   0.78   \\
         &   &   & 2003-04-03 & 332.8 &  47.09 &    0.61 &  74.03  &   0.79   \\
         &   &   & 2003-04-17 & 346.6 &  47.75 &    0.62 &  72.72  &   0.78   \\
         &   &   & 2003-06-21 & 411.2 &  49.48 &    0.63 &  73.19  &   0.78   \\
NGC~4736b& 300&  300& 2003-03-20 & 318.6 &  28.21 &    0.44 &  20.74  &   0.27   \\
         &   &   & 2003-04-03 & 332.8 &  29.94 &    0.45 &  21.32  &   0.27   \\
         &   &   & 2003-04-17 & 346.6 &  28.27 &    0.44 &  20.42  &   0.27   \\
         &   &   & 2003-06-21 & 411.2 &  29.64 &    0.45 &  21.50  &   0.28   \\
NGC~5055 & 300& 300& 2002-07-19 &  74.0 &  77.57 &    0.92 &  91.32  &   0.97   \\
         &   &   & 2003-03-12 & 310.3 &  77.27 &    0.92 &  89.56  &   0.95   \\
         &   &   & 2003-03-31 & 329.9 &  76.38 &    0.91 &  90.48  &   0.96   \\
         &   &   & 2003-04-22 & 351.1 &  75.66 &    0.90 &  91.24  &   0.97   \\
         &   &   & 2003-07-02 & 422.5 &  75.00 &    0.90 &  91.30  &   0.97   \\
NGC~6500 & 300& 300& 2002-07-06 &  61.0 &  31.49 &    0.62 &  27.11  &   0.37   \\
         &   &   & 2003-04-13 & 342.1 &  31.85 &    0.62 &  27.27  &   0.38   \\
         &   &   & 2003-04-20 & 349.9 &  32.51 &    0.63 &  26.63  &   0.37   \\
         &   &   & 2003-06-10 & 400.3 &  29.70 &    0.61 &  27.58  &   0.38  
\enddata
\tablecomments{(2)-(3)- Exposure time, in seconds. An empty entry
 indicates the same exposure time as above it; 
(5)- Modified Julian Date $- 2452400$; (6)-(9)- Nuclear flux densities and
  $1\sigma$ errors
  in units of $10^{-17}\ergcmsA$. NGC~4736b is the off-nuclear UV
  source in NGC~4736. See text for details of photometry
  and calibration.}
\end{deluxetable}

\begin{deluxetable}{lccrcrrrrrrrrrrr}
\tablecolumns{16}
\tabletypesize{\scriptsize}
\tablewidth{0pt}
\tablecaption{Measured Properties
\label{table2}}
\tablehead{
\colhead{Object} &
\colhead{type} &
\colhead{radio} &
\colhead{D} &
\colhead{Ref} &
\colhead{$\bar f_{\lambda}$} &
\colhead{$\bar f_{\lambda}$} &
\colhead{$\log L_{\lambda}$} &
\colhead{UV}& 
\colhead{n} &
\colhead{$\chi^2_{\rm dof}$} &
\colhead{$\Delta/\bar f$} &
\colhead{$\chi^2_{\rm dof}$} &
\colhead{$\Delta/\bar f$} &
\colhead{$\Delta$} &
\colhead{$\sigma$} \\
\colhead{} &
\colhead{} &
\colhead{} &
\colhead{Mpc} &
\colhead{} &
\colhead{F250W} &
\colhead{F330W} &
\colhead{F250W} &
\colhead{color} &
\colhead{} &
\colhead{F250W} &
\colhead{F250W} &
\colhead{F330W} &
\colhead{F330W} &
\colhead{color} &
\colhead{} \\
\colhead{(1)} &
\colhead{(2)} &
\colhead{(3)} &
\colhead{(4)} &
\colhead{(5)} &
\colhead{(6)} &
\colhead{(7)} &
\colhead{(8)} &
\colhead{(9)} &
\colhead{(10)} &
\colhead{(11)} &
\colhead{(12)} &
\colhead{(13)} &
\colhead{(14)} &
\colhead{(15)} &
\colhead{(16)} 
 }
\startdata
NGC~  404&2& N&    3.05&a,b&   73.57&   85.27&   36.11&     .97& 1&     \nodata&     \nodata&     \nodata&     \nodata&     \nodata&     \nodata \\
NGC~ 1052&1& Y&   18.03&a&    7.65&    8.97&   36.57&     .90& 1&     \nodata&     \nodata&     \nodata&     \nodata&     \nodata&     \nodata \\
M81      &1& Y&    3.63&a,e&  203.05&  130.25&   36.77&    1.83& 5&{\bf 9.28}&     .08&{\bf 18.36}&     .11&     .97&     .02 \\
NGC~ 3368&2& N&   10.67&a,f&   22.26&   29.81&   36.58&     .78& 2&    1.73&     .02&     .55&     .00&     .97&     .03 \\
NGC~ 3486&2& N&    7.40&g&   10.76&   18.15&   35.93&     .62& 2&     .20&     .00&     .15&     .00&    1.00&     .03 \\
NGC~ 3642&1& N&   27.50&g&   24.49&   22.13&   37.40&    1.13& 4&{\bf 3.22}&     .07&     .91&     .02&    1.04&     .03 \\
NGC~ 3998&1& Y&   13.14&a&  198.66&  152.64&   37.68&    1.34& 5&{\bf 41.01}&     .19&{\bf 51.13}&     .18&    1.00&     .03 \\
NGC~ 4203&1& Y&   15.14&a&   58.23&   37.19&   37.26&    1.60& 5&{\bf 263.09}&     .46&{\bf 160.50}&     .34&    1.11&     .02 \\
NGC~ 4258&?& Y&    7.30&a,c,i&    5.16&   11.12&   35.59&     .48& 5&{\bf 5.08}&     .15&{\bf 5.20}&     .07&    1.08&     .05 \\
M87      &1& Y&   15.42&a,h&   55.34&   48.42&   37.29&    1.19& 2&    3.52&     .03&     .35&     .00&    1.02&     .02 \\
NGC~ 4552&?& Y&   15.35&a&    1.97&    1.49&   35.89&    1.44& 2&{\bf 18.62}&     .18&{\bf 16.17}&     .19&     .99&     .06 \\
NGC~ 4569&2& N&   11.86&h&  992.54&  837.16&   38.38&    1.30& 4&     .60&     .01&     .24&     .00&    1.01&     .02 \\
NGC~ 4579&1& Y&   20.99&h&   61.11&   41.88&   37.65&    1.58& 2&{\bf 14.62}&     .07& {\bf 20.00}&     .07&     .99&     .02 \\
NGC~ 4594&2& Y&    9.08&a&    7.47&   15.29&   35.97&     .52& 3&{\bf 9.32}&     .19&{\bf 16.25}&     .11&    1.08&     .06 \\
NGC~ 4736&2& Y&    4.89&a,d&   47.99&   73.20&   36.21&     .68&4&{\bf 2.69}&     .05&     .56&     .01&    1.03&     .02 \\
NGC~4736b&?& N&    4.89&a,d&   29.01&   20.99&   35.99&    1.43& 4&{\bf 4.03}&     .06&{\bf 3.33}&     .05&    1.01&     .03 \\
NGC~ 5055&2& N&    7.40&g&   76.38&   90.78&   36.77&     .87& 5&    1.44&     .03&     .63&     .01&    1.01&     .02 \\
NGC~ 6500&2& Y&   39.70&g&   31.39&   27.15&   38.07&    1.38& 4& {\bf  3.92} &     .09&    1.10&     .03&    1.06&     .03 \\
\enddata
\tablecomments{(2)- Type-1 or type-2 object, depending on presence or
  absence, respectively, of broad H$\alpha$. NGC~4258 and NGC~4552 do
  not fall easily into either category (see text), and the spectral
  type of the off-nuclear source NGC~4736b is unknown. These three are
  marked with a ``?''; (3) - compact radio core detected (Y) or
  undetected (N); (4) - distance; (5) - distance reference (see below).
When several measurements
  exist for a galaxy, their average was adopted;
 (6)-(7) - flux density, averaged over all epochs, 
in units of $10^{-17}\ergcmsA$; (8) - log of monochromatic
 luminosity
at 2500~\AA, corrected for Galactic extinction, in
units of ergs~s$^{-1}$\AA$^{-1}$; (9) - UV color,  $f_{\lambda}$(F250W)/$f_{\lambda}$(F330W), after correction for Galactic
reddening (see text); (10) - number of epochs at which observations were made;
(11), (13) - $\chi^2$ per degree of freedom
compared to a constant at the mean level. Values implying variability
at $>95\%$ confidence are in boldface; (12), (14) - peak-to-peak
variation amplitude, with noise subtracted in quadrature, as a
fraction of mean flux; (15)-(16) - UV color change between  
the two epochs with extreme fluxes -- 
[$f_{\max}$(F250W)/$f_{max}$(F330W)]/
[$f_{min}$(F250W)/$f_{min}$(F330W)] -- and its uncertainty. 
}
\tablerefs{
 a - Tonry et
  al. (2001); b - Karachentsev et al. (2002); c - Newman et al. (2001); 
d - Karachentsev et al. (2003); e - Freedman et al. (2001); f - Tanvir et
al. (1999); g - Tully (1988); h - Gavazzi et al. (1999); 
i - Herrnstein et al. (1999).}

\end{deluxetable}

\end{document}